\documentclass[twocolumn,pre,showpacs,byrevtex]{revtex4-1}
\usepackage{epsf,graphicx,amssymb}
\usepackage[usenames]{color}
\usepackage{amsfonts}
\usepackage{amsmath}

\begin{document}
\title{Energy flux measurement from the dissipated energy in capillary wave turbulence}
\author{Luc Deike}
\author{Michael Berhanu}
\author{Eric Falcon}
\affiliation{Univ Paris Diderot, Sorbonne Paris Cit\'e, MSC, UMR 7057 CNRS, F-75 013 Paris, France, EU}

\date{\today}

\begin{abstract}We study experimentally the influence of dissipation on stationary capillary wave turbulence on the surface of a fluid by changing its viscosity.
We observe that the frequency power law scaling of the capillary spectrum departs significantly from its theoretical value when the dissipation is increased.  The energy dissipated by capillary waves is also measured and found to increase nonlinearly with the mean power injected within the fluid. Here, we propose an experimental estimation of the energy flux at every scale of the capillary cascade. The latter is found to be non constant through the scales. For fluids of low enough viscosity, we found that both capillary spectrum scalings with the frequency and the newly defined mean energy flux are in good agreement with wave turbulence theory. The Kolmogorov-Zakharov constant is then experimentally estimated and compared to its theoretical value.
\end{abstract}

\maketitle

\section{Introduction}
When a large ensemble of weakly nonlinear waves interact each other, they can develop a regime of wave turbulence where the wave energy is transferred from the large forcing scales to the small dissipative scales.  Exact solutions of out of equilibrium dynamics for the spectral content of energy can be derived analytically by a statistical theory called weak turbulence theory \cite{ZakharovBook,NazarenkoBook}. This theory can be applied in various contexts involving waves at various scales: astrophysical plasmas, internal waves in oceanography or in atmosphere, spin waves in magnetic materials, nonlinear waves in optics, etc. Because of hypotheses of weakly nonlinear waves, infinite system, local interactions and scale separations between energy source and dissipation, the applicability of weak turbulence to real systems can be questionable and experimental results are often in disagreement with the theory (see \cite{FalconReview,NewellReview} for recent reviews). In experiments, dissipation is often present at every scale and could explain some of these discrepancies. For instance, the spectrum of wave turbulence on an elastic plate has been experimentally shown to depart from its prediction when dissipation is increased \cite{Humbert2013}, whereas numerical works have shown that the theoretical spectrum is recovered when dissipation within the inertial range is removed \cite{Josserand,Miquel2013}.

Capillary waves are likely the easiest system to investigate wave turbulence in laboratory. Numerous experiments have then been dedicated to stationary capillary wave turbulence on the surface of fluid of low viscosity \cite{Falcon07a,Berhanu2013,Issenmann,Falcon0g,Holt1996,Putterman1996,Levinsen2000,Brazhnikov02,Xia10}. For capillary wave turbulence, weak turbulence theory predicts that the Kolmogorov-Zakharov spectrum of the wave height reads \cite{ZakharovBook}
\begin{equation}
S_{\eta}(f)=C^{KZ} \epsilon^{1/2}\left(\frac{\gamma}{\rho}\right)^{1/6}f^{-17/6},
\label{wtcap}
\end{equation}
where $\epsilon$ is the mean energy flux cascading through the scales, $\gamma$ the surface tension, $\rho$ the fluid density, $f$ the wave frequency and $C^{KZ}$ the Kolmogorov-Zakharov constant, that can be determined theoretically.  Such a frequency scaling $S_{\eta}\sim f^{-17/6}$ has been observed either numerically \cite{Pushkarev1996,Pushkarev2000}, or experimentally using vibrating plunging wave makers \cite{Falcon07a,Berhanu2013}, vibrating the whole container \cite{Issenmann}, and working in low-gravity \cite{Falcon0g} or acoustically levitated \cite{Holt1996} environments.  Note that with parametric forcing, peaks and forcing harmonics are observed on the spectrum with maximal amplitudes decreasing roughly as $f^{-17/6}$ \cite{Putterman1996,Levinsen2000,Brazhnikov02,Xia10}.

Nevertheless, some questions still remain open. For instance, experiments show a spectrum scaling with the energy flux in disagreement with the one predicted by the theory \cite{Falcon07a,Xia10,Issenmann}. The energy flux is usually estimated by measuring the injected power in the fluid that is assumed to be transferred in the wave system without dissipation within the inertial range. Another attempt to estimate the mean energy flux consists of measuring the wave energy decay rate after switching off the wave maker \cite{Denissenko07}. However, in a precedent paper \cite{Deike2012}, we have experimentally shown that the energy decay in gravity-capillary wave turbulence is mainly piloted by large scale viscous dissipation.

In this paper, we will study stationary gravity-capillary wave turbulence on the surface of fluids of different viscosities. We show that the frequency scaling of the capillary spectrum departs from its theoretical prediction when dissipation is increased. By measuring the injected power in the fluid, together with the dissipated powers by gravity and capillary waves, we show that most of the injected energy is dissipated at large scales by gravity waves, whereas a small part feeds the capillary cascade. Moreover, the energy dissipated by capillary waves is found to increase nonlinearly with the mean injected power. Both results mean that estimating the mean energy flux in the capillary cascade by the injected power is not valid. Here, we propose an original estimation of the energy flux at every scale of the capillary cascade from the experimental energy spectrum and the wave dissipation rate. This energy flux is then found to be non constant through the capillary scales contrary to the assumptions. However, defining a mean energy flux over the scales allow us to rescale the wave spectrum with the mean energy flux in good agreement with wave turbulence theory for fluids of low enough viscosity. The Kolmogorov-Zakharov constant is then evaluated experimentally, for the first time.

The paper is organized as follows. In Sect. \ref{theory}, we recall the origin on wave dissipation in wave turbulence on the surface of a fluid. The experimental setup is described in Sect. \ref{expset}. The experimental results are then discussed: the evolution of the wave spectrum when the dissipation is increased (Sect. \ref{part1}), the measurement of the dissipated powers by gravity and capillary waves, and their corresponding spectra (Sect. \ref{part2}). Finally, we present the experimental estimation of the energy flux at every scale (Sect. \ref{part3}) and of the Kolmogorov-Zakharov constant (Sect. \ref{part4}). A conclusion is given in Sect. \ref{Conclusion}.

\section{Origin of wave dissipation\label{theory}}
Dissipation of propagating waves in a closed basin has been studied theoretically and experimentally by various authors \cite{Miles,LandauFluid,Lamb}. Linear viscous dissipation leads to an exponential decay of the wave: $\eta(t)=\eta_0 e^{-\Gamma t}$, with $\eta_0$ the initial amplitude of the wave, and $\Gamma^{-1}$ its theoretical damping time that depends on the frequency and the nature of dissipation. Wave damping can have different origins:  bottom boundary layer ($\Gamma_{B}$), side wall boundary layer ($\Gamma_{W}$), and surface dissipation. Two types of surface dissipation can be considered: the classical viscous dissipation at a free surface $\Gamma_{\nu} \sim \nu k^2$ \cite{LandauFluid,Lamb}, or viscous dissipation in presence of an inextensible film $\Gamma_{S} \sim (\nu f)^{1/2}k$ \cite{Miles,Lamb}. The latter comes from the presence of surfactants/contaminants at the interface that leads to an inextensible surface where the tangential velocity is cancelled at the interface and was first considered to study the effect of the calming effect oil on water. Note that these surface dissipations are incompatible since they correspond to two different kinematic conditions at the interface \cite{Lamb}. The decay rate for the wave of frequency $f$ is defined by $\delta\equiv \Gamma/(2\pi f)$. The theoretical decay rate for the various types of viscous dissipation in a fluid of arbitrary depth $h$ are \cite{Miles,LandauFluid,Lamb}

\begin{align}
\delta_{\nu} & =\frac{\nu k^{2}}{\pi f} \label{eqnu}\\
\delta_{S} & =\left(\frac{\nu}{4\pi f}\right)^{1/2}\frac{k\cosh^{2}{kh}}{\sinh{2kh}} \label{eqS}\\
\delta_{B} & =\left(\frac{\nu}{4\pi f}\right)^{1/2}\frac{k}{\sinh{2kh}} \label{eqB}\\
\delta_{W} & =\left(\frac{\nu}{4\pi f}\right)^{1/2}\frac{1}{2R}\left(\frac{1+(m/kR)}{1-(m/kR)}-\frac{2kh}{\sinh{2kh}}\right) \label{eqW}
\end{align}
where $R$ is the size of the circular vessel, and $m=1$ the anti-symmetrical modes and $m=0$ the symmetrical ones. 

In a precedent paper \cite{Deike2012}, we have experimentally shown that the major part of dissipation occurs at large scales in gravity-capillary wave turbulence, and that the experimental decay rate scales as $\nu^{1/2}$ over two decades in viscosity, and not as $\nu^1$ as expected by the classical viscous dissipation. In our experiments, viscous dissipations by surface boundary layer and bottom boundary layer are the most important while friction at the lateral boundary is negligible \cite{Deike2012}. Bottom friction is significant at large scale since the forcing scales are of the order of the depth. The experimental wave dissipation is correctly described by the total theoretical dissipation: 
\begin{equation}
\Gamma(f)=2\pi f\delta_{T}=2\pi f(\delta_{S}+\delta_{B}+\delta_{W}).
\label{disseq}
\end{equation}
The fact that the inextensible condition has to be taken into account instead of the usual free surface condition was previously observed in laboratory experiments with water \cite{Miles,VanDorn,HendersonMiles}. Indeed, if no particular attention is paid (such as working in clean room, filtered fluid or fluid with low enough surface tension), the surface dissipation by boundary layer dominates the $\nu k^2$ dissipation \cite{HendersonMiles}. Finally, note that the infinite depth condition is satisfied for $f>10$ Hz (i.e. $\lambda <2$ cm and $kh \gg 1$), and thus bottom friction becomes also negligible for capillary waves. Consequently, in our experiments, the dissipation source for capillary waves is only due to surface dissipation in presence of an inextensible film \cite{Deike2012}. In the following, we will estimate the damping rate using the full Eq.\ (\ref{disseq}) since gravity and capillary waves are involved in our experiments. 

\begin{figure}[t!]
\begin{center}
\includegraphics[scale=1.2]{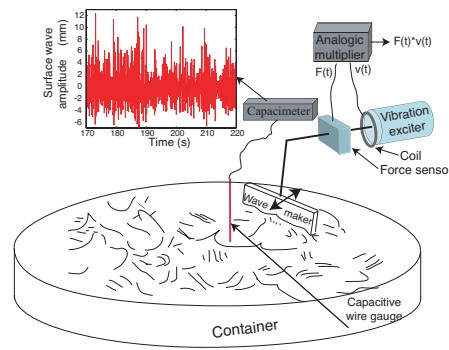}
\end{center}
\caption{(Color online) Experimental setup. The diameter of the vessel is $22$ cm.}
\label{dispo_stat}
\end{figure}

\section{Experimental setup\label{expset}}
The experimental setup is the same as in the experiment on freely decaying wave turbulence \cite{Deike2012} and similar to the one used in~\cite{Falcon07a}. It consists of a circular plastic vessel, 22 cm in diameter, filled with a fluid to a height $h=25$ mm. Various fluids are used: water, mercury, silicon oils, and aqueous solutions of glycerol (denoted as x\%GW with x the glycerol percent) to vary kinematic viscosity, $\nu$, over two orders of magnitude. Properties of these fluids are listed in Table \ref{tab}. The main difference between the different fluids is their kinematic viscosity. The theoretical gravity-capillary transition $f_{gc}=\frac{1}{2\pi}\sqrt{2g/l_c}$ is between 14 and 19 Hz in all cases, with $l_c=\sqrt{\gamma/(\rho g)}$ the capillary length, $\rho$ the density, and $\gamma$ the surface tension. 

\begin{table}[t]
\begin{center}
\begin{tabular}{lcccc}
\hline
\hline
Fluid & $\rho$ (kg/m$^{3}$) & $\nu$ (m$^{2}$/s) & $\gamma$ (mN/m) & $f_{gc}$ (Hz)\\
\hline
Mercury & 13 600 & 1.1$\ $10$^{-7}$ & 400 & $17$\\
Water & 1 000 & 10$^{-6}$ & 73 & $14$\\
20\% Glycerol-Water & 1 020 & 2$\ $10$^{-6}$ & 70 & $13.5$\\
30\% Glycerol-Water & 1 050 & 3$\ $10$^{-6}$ & 70 & $14$\\
50\% Glycerol-Water & 1 120 & 5$\ $10$^{-6}$ & 68 & $14$\\
Silicon oil V5 & 1 000 & 5$\ $10$^{-6}$ & 20 & $18.8$\\
Silicon oil V10 & 1 000 & 10$^{-5}$ & 20 & $18.5$\\
\hline
\hline
\end{tabular}
\caption{Physical fluid properties: density, $\rho$, kinematic viscosity, $\nu$, and surface tension $\gamma$ \cite{Glycerin}. The frequency transition between gravity and capillary waves is $f_{gc}$ (see text).}
\label{tab}
\end{center}
\end{table}

Surface waves are generated by a rectangular plunging wave maker (13~cm in length and 3.5 cm in height) driven by an electromagnetic vibration exciter (LDS V406) driven by a random noise (in amplitude and frequency) band-pass filtered typically between 0.1 and 5 Hz. The wavemaker is continuously driven and the wave height $\eta(t)$ is recorded during the stationary regime (300 s acquisition time) at a given location (center of the vessel) by a capacitive wire gauge plunging perpendicularly to the fluid at rest~\cite{Deike2012,Falcon07a}. The capacitive gauge is calibrated for each fluid and we have checked than the response is linear with the wave height whatever the working fluid.

The force $F(t)$ applied by the shaker to the wavemaker and the velocity $V(t)$ of the wavemaker are measured to access to the injected power $I=F \times V$ into the fluid \cite{Falcon07a}. We have checked that the classical relation $\langle I \rangle \sim \rho \sigma_{V}^2$ \cite{Falcon07a,FalconFlux} between the wave maker rms velocity and the mean injected power holds for every fluid. Moreover, a scaling $\langle I \rangle\sim \nu^{1/2}$ is observed, compatible with the observed dissipation. One observes also $\langle I \rangle/\rho \sim \sigma_{\eta}^2$ for all fluids, which is due to the fact that the gravity wave energy scales as $\sim g\eta^2$. The mean injected power value is thus directly related to the rms wave height.  $\langle I \rangle$ is normalized by $\rho$ and the vessel surface $S=\pi R^2$ to compare the results without considering the inertial effects
\begin{equation}
\epsilon_I=\frac{\langle I \rangle}{\rho S}\ .
\end{equation}
$\epsilon_I$ has thus the dimension of an energy flux by density unit ($[L^3 T^{-3}]$), as for the theoretical mean energy flux $\epsilon$.

\begin{figure}
\begin{center}
\includegraphics[scale=0.45]{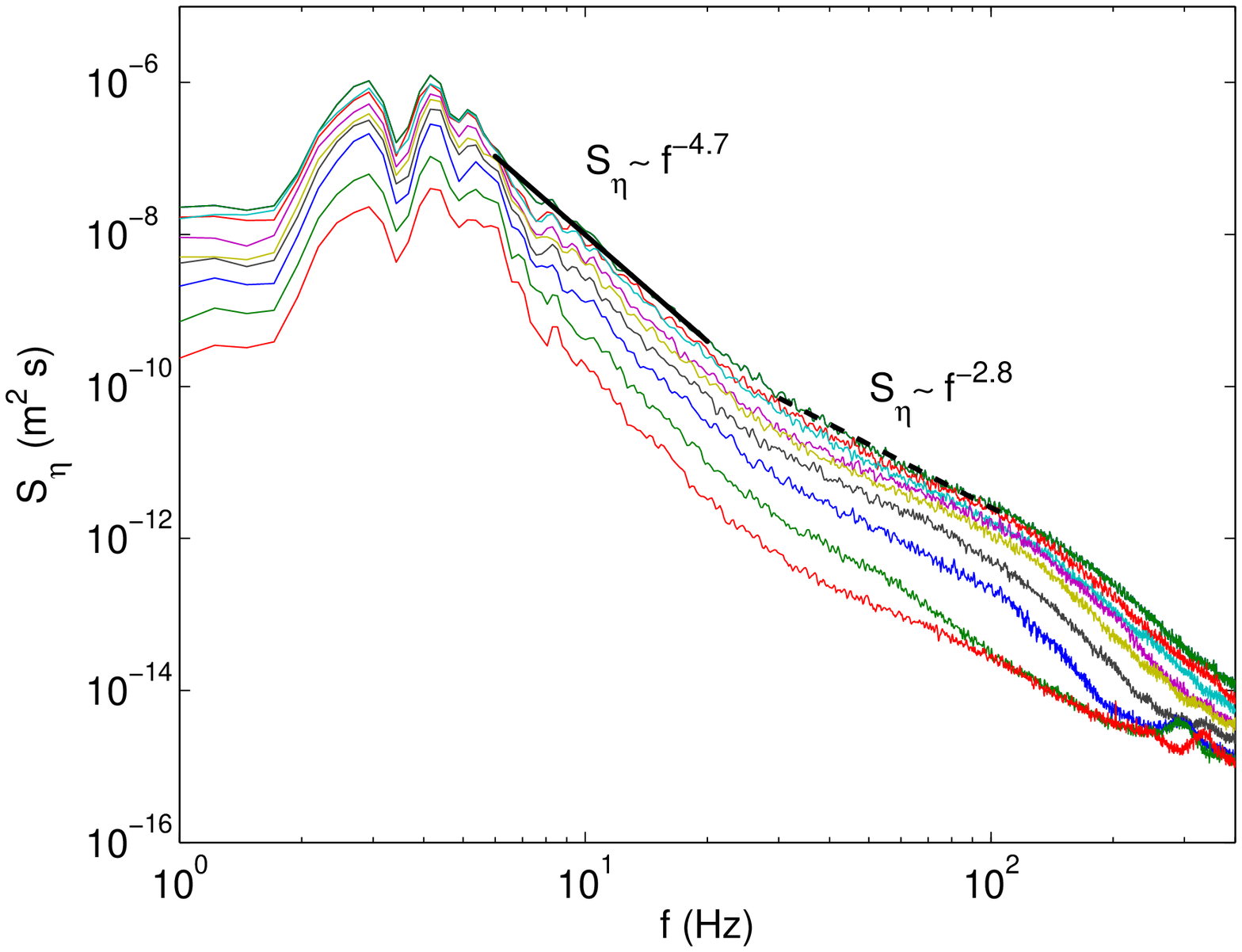}\\
\includegraphics[scale=0.45]{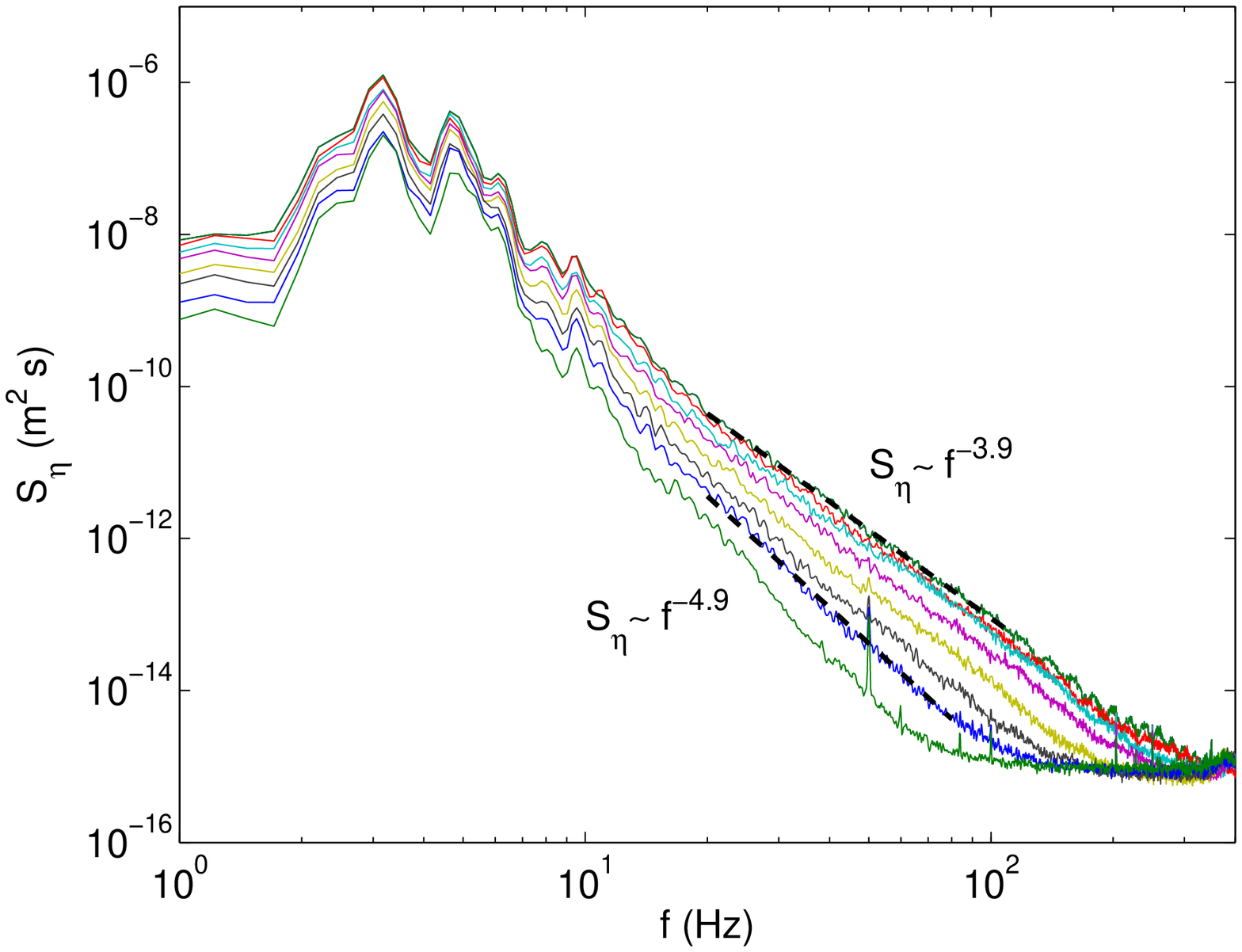}\\
\includegraphics[scale=0.45]{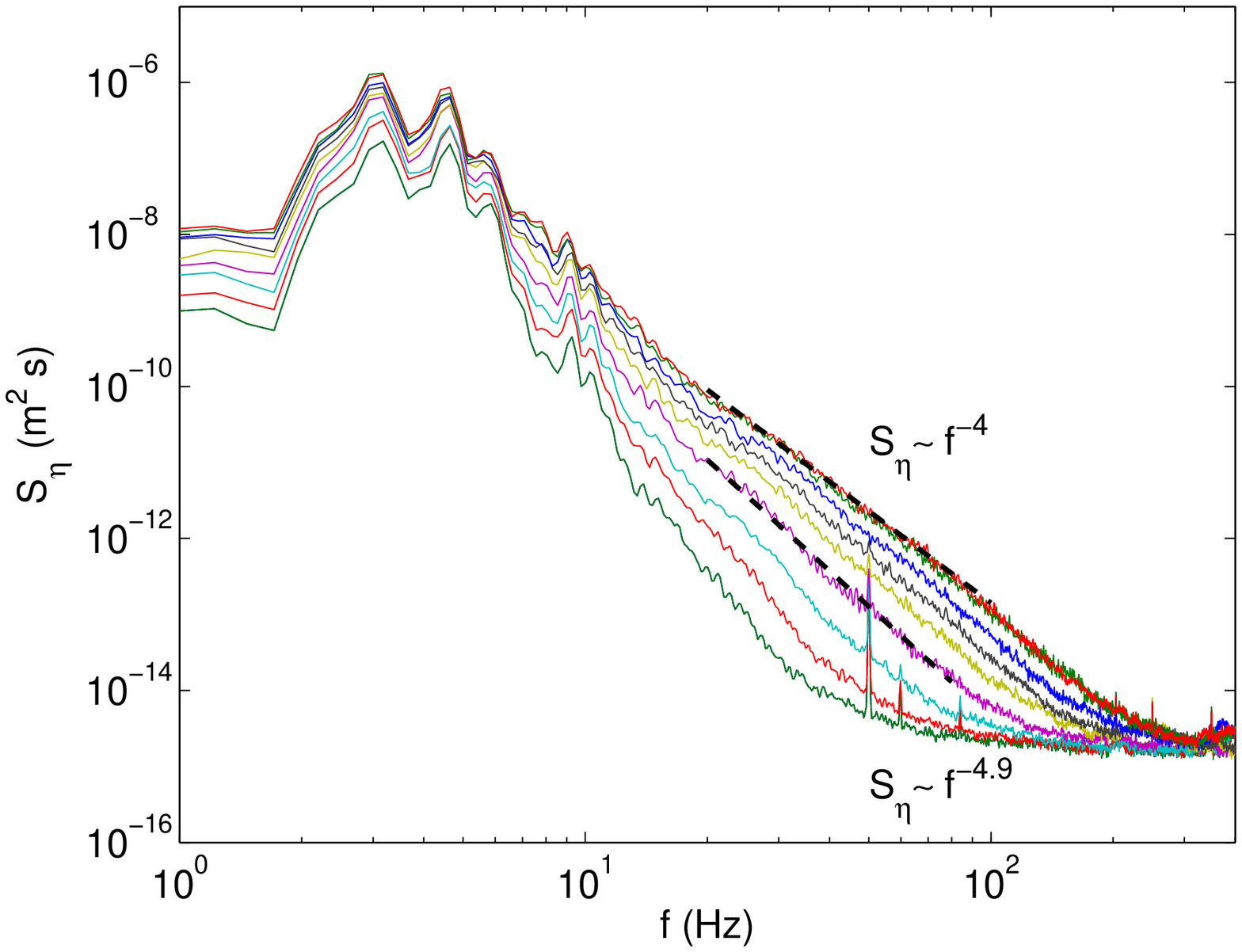}
\caption{(Color online) Power spectrum $S_{\eta}(f)$ for low (top), medium (middle) and high (bottom) viscosity corresponding respectively to mercury, 30\% GW, and 50\% GW of viscosity $\nu=1.1\ 10^{-7}$, $3\ 10^{-6}$, and $5\ 10^{-6}$  m$^2$/s. Injected power increases from bottom to top. Dashed lines are power law fits.}
\label{spvisc1}
\end{center}
\end{figure}

\section{Role of dissipation in capillary wave turburlence\label{part1}}
In this section, we investigate the influence of an increasing dissipation on capillary wave turbulence.

\subsection{Power spectrum of wave height}
We first focus on the power spectrum of wave height, $S_{\eta}(f)$, on the surface of various fluids of different viscosities. Figure \ref{spvisc1} shows $S_{\eta}(f)$ for different viscosities ($1.1\ 10^{-7}\leq \nu \leq 5\times 10^{-5}$ m$^2$/s), and different forcing amplitudes.

For {\em low dissipation} (i.e. small viscosity as in mercury or water), $S_{\eta}(f)$ displays two frequency power laws whatever the forcing amplitude [see in Fig. \ref{spvisc1}(a)], corresponding to the gravity (6 Hz $< f < f_{gc}$) and capillary ($f_{gc}< f\lesssim 120$ Hz) wave turbulence regimes. The transition between these two regimes is observed around the theoretical gravity-capillary transition frequency $f_{gc}$. The gravity spectrum is found as $S_{\eta}^g\sim f^{-\beta}$, with $4. 5\leq \beta \leq 5.5$, steeper than the theoretical spectrum ($\sim f^{-4}$) and depends on the injected power. Within the capillary inertial range ($f_{gc} < f \lesssim 120$ Hz), one has $S_{\eta}\sim f^{-\alpha}$, with $\alpha=2.8\pm0.2$ independent of the injected power, and in good agreement with wave turbulence theory ($\sim f^{-17/6}$). At higher frequencies ($f>f_d\approx120$ Hz), the spectrum shape changes to due an increase of dissipation. All these results are similar to those found in \cite{Falcon07a,Deike2012}.

For higher viscosities ($\nu > 2\times 10^{-6}$ m$^2$/s), the spectrum phenomenology changes as shown in Figs. \ref{spvisc1}(b-c). It is not possible anymore to define a cascade within the gravity wave range, the power law has been replaced by peaks, corresponding to the vessel eigenvalues and their harmonics. However, a power law is still observed in the capillary wave range, $S_{\eta}\sim f^{-\alpha}$, with $\alpha$ larger than its theoretical value and dependent on the injected power: The wave spectrum is steeper when the injected power decreases. These observations are valid for all considered fluids with $\nu > 2\times 10^{-6}$ m$^2$/s, both in aqueous solutions of glycerol and in silicon oil. We will refer below this behavior as the {\it{high dissipation}} regime of wave turbulence. Finally, when the viscosity is increased, a change of curvature of spectrum shapes is observed near high frequencies ($f\gtrsim 120$ Hz) in Fig.\ \ref{spvisc1}. For high enough viscosity, the capillary cascade gets directly into the noise level, which can be ascribed to the lower sensitivity of the capacitive gauge when the glycerol concentration is increased.

\begin{figure}
\begin{center}
\includegraphics[scale=0.45]{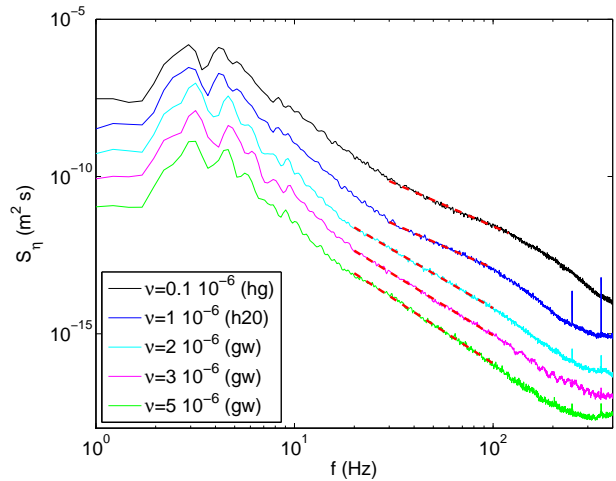}
\caption{(Color online) $S_{\eta}(f)$ for various fluids: mercury, water, 20\% GW, 30\% GW, and 50\% GW ($\nu=1.1\ 10^{-7}$, $10^{-6}$, $2\ 10^{-6}$, $3\ 10^{-6}$, and $5\ 10^{-6}$ m$^2$/s (from top to bottom). $\epsilon_I\approx 5\ 10^{-5}$ m$^3$s$^{-3}$. Curves are shifted vertically for clarity by a factor 1, 0.5, 0.1, 0.01, 0.001 respectively. Dotted (red) lines show best power law fits, $S_{\eta}\sim f^{-\alpha}$, with $\alpha=$2.8, 2.8, 3.7, 3.9, and 4.1 (from top to bottom).}
\label{spbilan}
\end{center}
\end{figure}
\begin{figure}
\begin{center}
\includegraphics[scale=0.45]{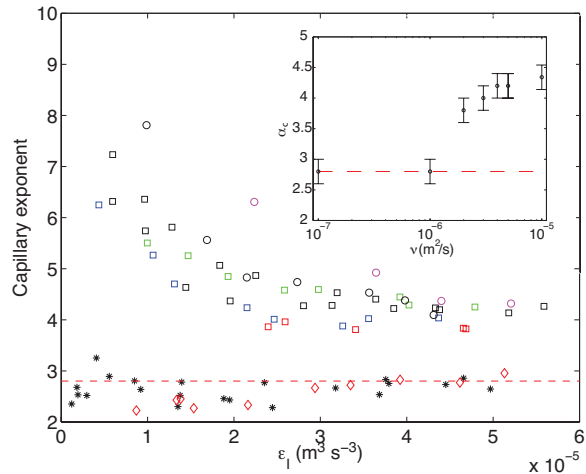}
\caption{(Color online) Main: Capillary exponent $\alpha$ as a function of $\epsilon_I$ for various fluids $\nu=1.1\ 10^{-7}$ ($\star$), $10^{-6}$  ({\color{red}$\diamond$}), $2\ 10^{-6}$ ({\color{red}$\square$}), $3\ 10^{-6}$ ({\color{blue}$\square$}), $4\ 10^{-6}$ ({\color{green}$\square$}), $5\ 10^{-6}$ (GW) ({\color{black}$\square$}), $5\ 10^{-6}$ (oil) ({\color{black}$\circ$}), and $10^{-5}$ m$^2$/s ({\color{magenta}$\circ$}). Inset: $\alpha$ vs. $\nu$ for fixed forcing $\epsilon_I\approx 5\ 10^{-5}$ m$^3$s$^{-3}$. The theoretical capillary exponent $\alpha=17/6$ is indicated in dashed (red) lines.\label{spbilan2}}
\end{center}
\end{figure}

\subsection{Frequency power-law exponent of the spectrum}
Figure \ref{spbilan} shows $S_{\eta}(f)$ at different kinematic viscosities for a fixed strong forcing. For the two lowest viscosity fluids, the spectrum exhibits two frequency power laws, corresponding to the gravity wave cascade, $S_\eta(f)\sim f^{-5\pm0.5}$, and the capillary one $S_\eta(f)\sim f^{-2.8}$. Thus at {\it{low dissipation}}, the capillary exponent is in good agreement with the wave turbulence prediction. When the dissipation is increased, a capillary cascade is still observed $S_\eta(f)\sim f^{-\alpha}$ but with an exponent $\alpha$ dependent on the viscosity as shown in the inset of Fig. \ref{spbilan2}.

Figure \ref{spbilan2} shows the capillary exponent $\alpha$ as a function of $\epsilon_I$. At low viscosity ($\nu \leq10^{-6}$ m$^2$/s), the exponent $\alpha=2.8\pm 0.2$, independent of $\epsilon_I$, as expected by the theory. At higher viscosity ($\nu \geq 2\ 10^{-6}$ m$^2$/s), $\alpha$ is larger than the theoretical value and depends on the injected power: $\alpha$ decreases with $\epsilon_I$ up to a saturating value at large forcing ($\epsilon_I > 3 \ 10^{-5}$ m$^3$s$^{-3}$). 

\subsection{Discussion}
The capillary cascade displays two qualitative behaviors regarding the amount of dissipation. When the dissipation is low enough, the theoretical scaling in frequency is observed and is independent of the injected power, as previously reported. When the dissipation is increased beyond a certain point, a steeper power law spectrum is observed. This discrepancy between theory and experiment becomes larger when the dissipation is further amplified. This result is very similar to the one recently reported in flexural wave turbulence \cite{Humbert2013}. Moreover, the frequency exponent of the wave spectrum power law depends on the injected power. This latter reminds us os what is observed in gravity wave turbulence \cite{Falcon07a,Denissenko07,Nazarenko2010}. Recent results in hydroelastic wave turbulence on the surface of a floating elastic sheet \cite{Deike2013} also shows a wave turbulence regime with a power law steeper than the one given by theoretical predictions. Dissipation could be also responsible of this dependency in those systems.

\section{Experimental determination of dissipated power by the waves\label{part2}}
The part of the injected power linearly dissipated by the waves will now be determined experimentally, using the experimental wave height spectrum $S_{\eta}(f)$ and the theoretical dissipation rate $\Gamma(f)$, and will be compared to the mean injected power at the wave maker $\epsilon_I$.

\subsection{Definitions}
The potential wave energy, per surface and density unit is $E_g=\frac{1}{2}g\eta^2$ for gravity waves and by $E_c=\frac{1}{2}\frac{\gamma}{\rho}k^2 \eta^2$ for capillary waves. For linear waves, the total energy is given by the sum of the kinetic and the potential terms, and both values are equal in average. Since, we do not measure the kinetic energy, the potential energy is multiplied by 2 to take into account the kinetic energy. The wave energy spectrum in the Fourier space $E_{f}$ is related to the total energy $E=\int E_f df$ where $E_f=E_{f}^g+E_{f}^c$, and to the wave height power spectrum $S_{\eta}(f)$ by
\begin{align}
E_{f}^g(f)&=gS_{\eta}(f) \text{,\ for gravity waves,} \label{grav}\\
E_{f}^c(f)&=\frac{\gamma}{\rho}k^2S_{\eta}(f) \text{,\ for capillary waves.} \label{capi}
\end{align}
We define the wave dissipation spectrum $D_{\eta}(f)$ by
\begin{equation}
D_\eta(f)=E_f(f) \Gamma(f),
\label{Deta}
\end{equation}
where $E_f(f)$ is the wave energy spectrum and $\Gamma=1/\mathcal{T}$ the theoretical dissipation rate of Eq.\ (\ref{disseq}). $D_{\eta}(f)$ can be split in two terms, the capillary wave dissipation spectrum and the gravity one, with $D_{\eta}(f)=D_\eta^g(f)+D_\eta^c(f)$ and:
\begin{align}
D_\eta^g(f)=&gS_{\eta}(f)\Gamma (f), \label{Detag}\\
D_\eta^c(f)=&\frac{\gamma}{\rho}k^2S_{\eta}(f)\Gamma (f). \label{Detac}
\end{align}
The total power dissipated linearly by the waves is then given by integrating the dissipation spectrum: 
\begin{equation}
D=\int D_\eta(f)df=\int E_{f}(f)\Gamma (f)df, \label{D}
\end{equation}
The capillary and gravity dissipated powers are separately calculated
\begin{align}
D_g=&\int_{f_{T}}^{f_{gc}} gS_{\eta}(f)\Gamma (f)df, \label{Dgrav} \\
D_c=&\int_{f_{gc}}^{f_s/2} \frac{\gamma}{\rho}k^2S_{\eta}(f)\Gamma (f)df, \label{Dcap}
\end{align}
the integration ranges are given by $f_T=1/T$ the lowest accessible frequency where $T=300$ s is the total measurement time, $f_{gc}$ the gravity-capillary transition, and $f_s$ the sampling frequency ($f_s=1$ kHz). The total power dissipated b the waves is given by $D=D_c+D_g$. Thus, if all the injected power by the wave maker goes into the waves, we should have the power budget
\begin{equation}
\epsilon_I\equiv\frac{\langle I\rangle}{\rho S} =D=D_g+D_c.
\label{bilanPID}
\end{equation}
The dimension of $\epsilon_I$, $D$, $D_c$, and $D_g$ is $[L^3T^{-3}]$, the same as the one of the energy flux of wave turbulence theory. Note that this power budget does not take into account wave dissipation by non linear processes or bulk dissipation (the fluid being supposed to be almost irrotational, all the dissipation takes place near the boundaries).

\begin{figure}[t]
\begin{center}
\includegraphics[scale=0.45]{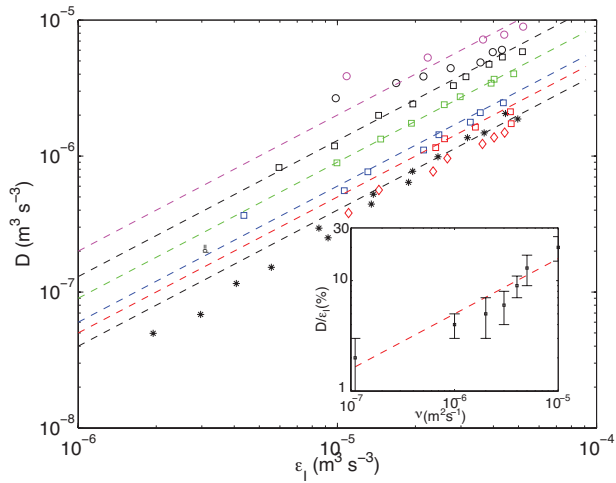}
\caption{(Color online) Main: Power dissipated by the waves $D$ as a function of $\epsilon_I$ for different fluids ($\nu$ increases from bottom to top). Symbols are the same as in Fig. \ref{spbilan2}. Dashed lines are linear fits $D=p(\nu) \epsilon_I$.  Inset: Part of the injected power dissipated in waves $p=D/\epsilon_I$ (in $\%$) as a function of $\nu$. Dashed line is the best fit $p\sim\nu^{1/2}$.}
\label{bilanpuiss1}
\end{center}
\end{figure}

\begin{figure}[t]
\begin{center}
\includegraphics[scale=0.45]{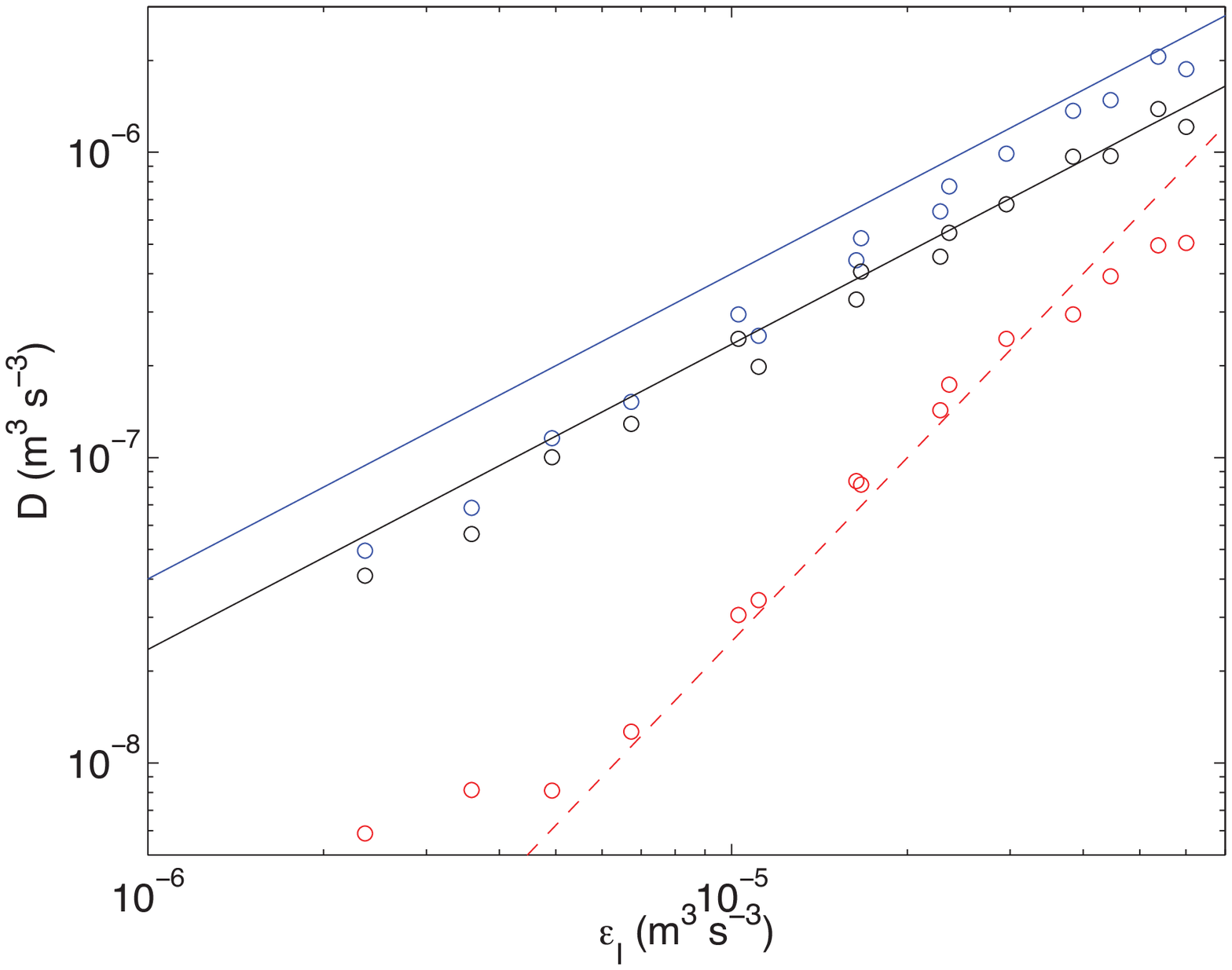}\\
\includegraphics[scale=0.45]{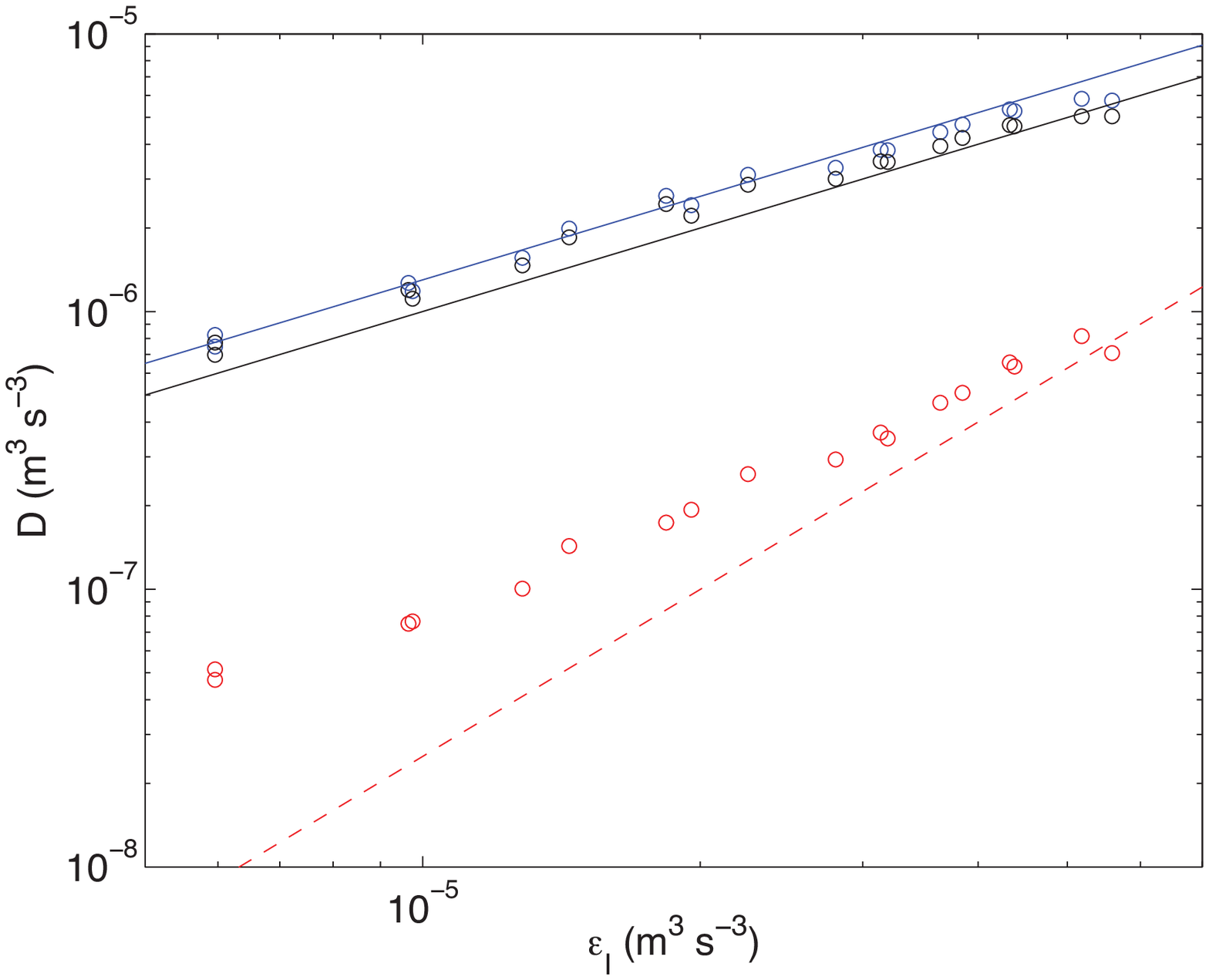}
\caption{(Color online) Power dissipated by waves, ($\circ$) $D$ (blue), $D_c$ (red) and $D_g$ (black) (from top to bottom) as a function of $\epsilon_I$, for mercury (top) and GW 50$\%$ (bottom). Solid lines are linear fits. Dashed lines show the fit $D_c\sim \epsilon_I^2$. Most of wave dissipation is done by gravity waves.}
\label{bilanpuiss2}
\end{center}
\end{figure}

\subsection{Dissipated power by the waves}
We first measure the total power dissipated by the waves, $D$, from the experimental power spectrum, $S_{\eta}(f)$, and using Eqs.\ (\ref{D}), (\ref{grav}), and  (\ref{capi}). Figure \ref{bilanpuiss1} shows that $D$ increases roughly linearly with $\epsilon_I$ for all fluids with a slope $p$ that depends on $\nu$. The inset of Fig. \ref{bilanpuiss1} shows that $p\sim \nu^{1/2}$ as expected by the definition $D \sim \Gamma$ [see Eq.\ (\ref{D})] and by the nature of dissipation $\Gamma \sim \nu^{1/2}$ (see \S \ref{theory}). To sum up, Fig. \ref{bilanpuiss1} shows that the power dissipated linearly by the waves is proportional to the mean injected power $D\sim \epsilon_I$. However, only a small part of the injected power is linearly dissipated by the waves. Indeed, the inset of Fig. \ref{bilanpuiss1}, shows that $p= D/\epsilon_I$ is only around $5\%$ in mercury, grows to $\approx 13\%$ in the GW solutions, and up to around $20\%$ in silicon oils. We will discuss later the possible mechanisms responsible for these observations. 

The dissipated power budget is shown in Fig. \ref{bilanpuiss2} in the case of low dissipation (mercury) and high dissipation (GW 50 \%). The total power dissipated $D$, the parts dissipated by gravity waves, $D_g$, and by capillary waves, $D_c$, are computed from the experimental power spectrum, $S_{\eta}(f)$, and using Eqs.\ (\ref{D}), (\ref{Dgrav}), and  (\ref{Dcap}). In both dissipation cases, the power dissipated by gravity waves is much larger than the one by capillary waves. Moreover, $D_g$ is roughly linear with $\epsilon_I$ whereas $D_c$ is found to scale nonlinearly with $\epsilon_I$ (e.g. $\sim \epsilon_I^2$  for mercury at high $\epsilon_I$). As explain below, this result will be of prime interest to understand the scaling of $S_{\eta}(f)$ with the energy flux. 

Calculating the ratio $D_g/D$ as a function of $\epsilon_I$ shows that from 85 $\%$ to 65 $\%$ of the wave dissipated power is dissipated by the gravity waves for mercury and $95\%$ to $85\%$ for GW fluids. Thus, not more than 35$\%$ of the dissipated power is due to capillary waves for mercury and less than 15 $\%$ for GW fluids. 

\begin{figure}[t]
\begin{center}
\includegraphics[scale=0.45]{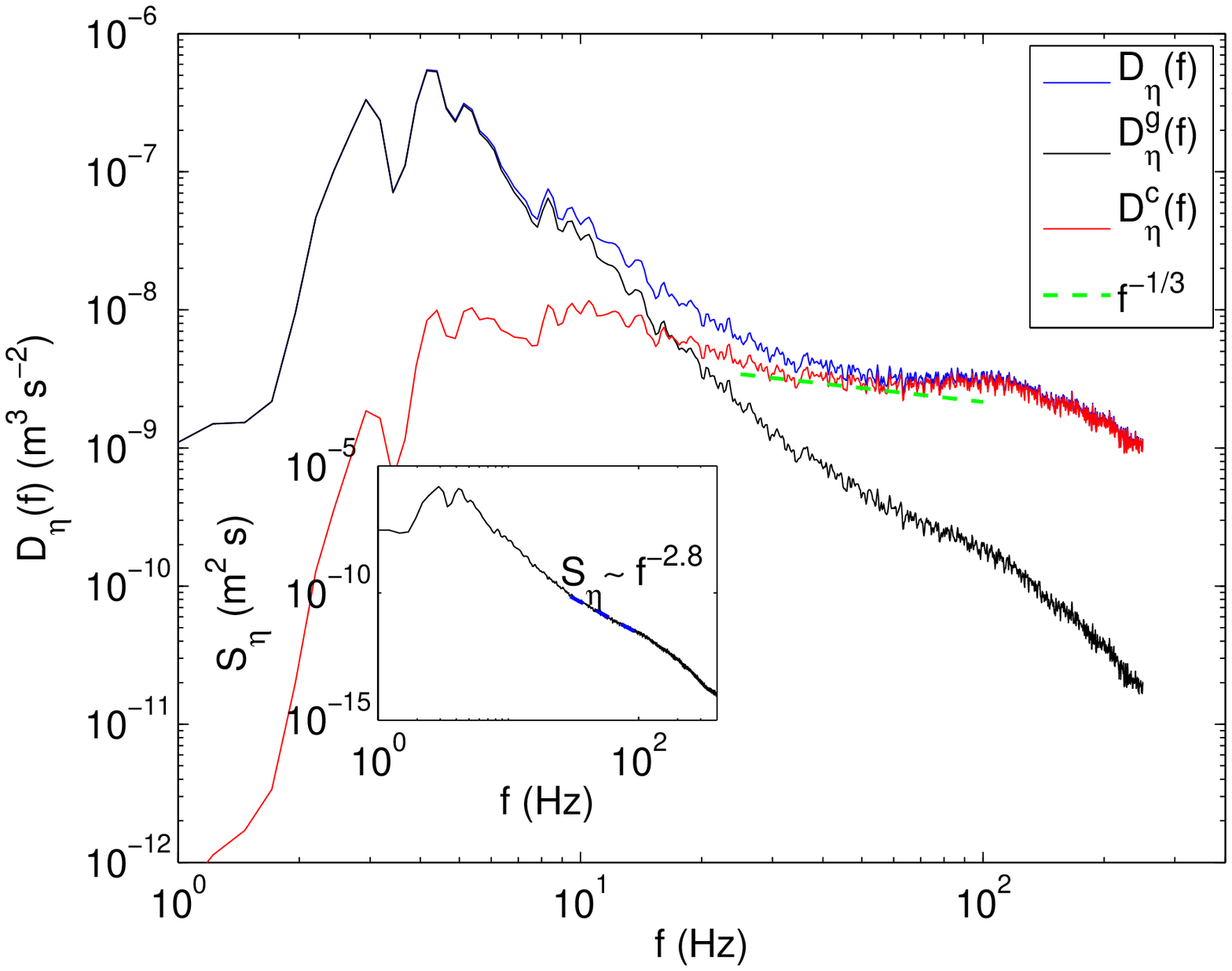}\\
\includegraphics[scale=0.45]{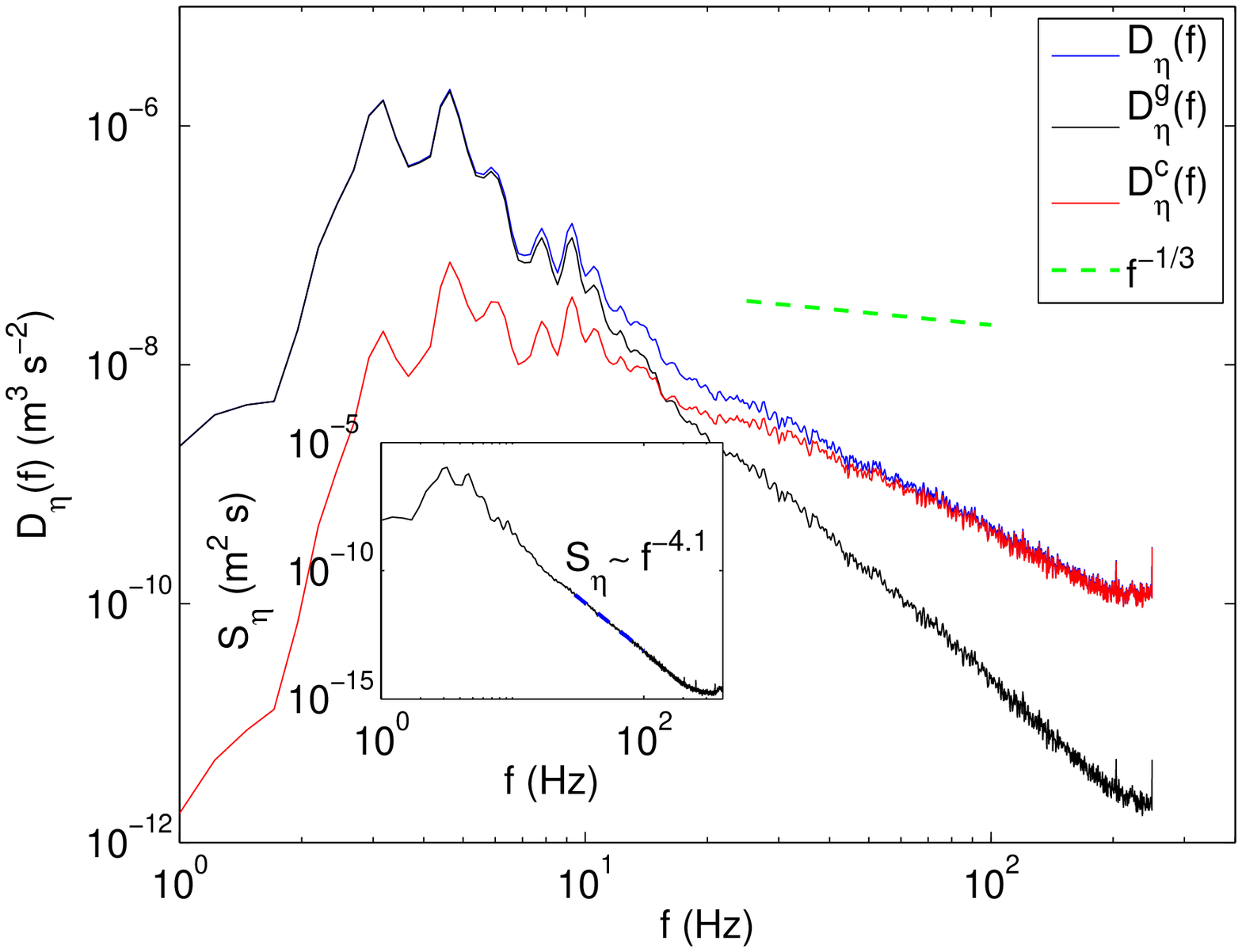}
\caption{(Color online) Spectrum of wave dissipation.  Mercury (top), 50$\%$ GW fluid (bottom). Solid lines (from top to bottom): dissipation spectra of all waves $D_{\eta}(f)$ (blue), of pure gravity waves $D_{\eta}^g(f)$ (black); and of pure capillary waves $D_{\eta}^c(f)$ (red). Dashed lines shows theoretical scaling $D_{\eta}^c \sim f^{-1/3}$. Inset: corresponding spectrum of wave height $S_{\eta}(f)$, where $f^{-2.8}$ (top) and $f^{-4.1}$ (bottom) are the best fits in the capillary range.}
\label{spectrediss}
\end{center}
\end{figure}

\subsection{Wave dissipation spectrum}
The spectrum of wave dissipation, $D_{\eta}(f)$, is obtained from the experimental power spectrum of wave height, $S_{\eta}(f)$, and using Eqs.\ (\ref{Deta}), (\ref{capi}), (\ref{grav}), and (\ref{disseq}). Figure \ref{spectrediss} shows $D_{\eta}(f)$ as a function of frequency, in the case of low dissipation (mercury) and high dissipation (GW 50 \%). The dissipation spectrum of gravity waves, $D^g_{\eta}(f)$, and of capillary waves, $D^c_{\eta}(f)$ are computed from $S_{\eta}(f)$, and using Eqs.\ (\ref{Detag}), (\ref{Detac}), and (\ref{disseq}). In both dissipation cases, Fig. \ref{spectrediss} shows that most dissipation occurs at large scales within the gravity wave frequency range, near the forcing scales. $D^g_{\eta}$ declines then abruptly at higher frequency. Note that the shape of $D^c_{\eta}$ is very different in the case of low and high dissipation. For low dissipation, a capillary cascade is observed in good agreement with wave turbulence theory (see inset of Fig. \ref{spectrediss} (top), and $D^c_{\eta}$ remains large at all scales:  $D^c_{\eta}$ is almost constant within the capillary inertial range ($f_{gc}\lesssim f\lesssim f_d\approx120$ Hz), before to slightly increases ($f_d\approx 120$ Hz), and to decreases abruptly after the end of the capillary cascade. For high dissipation, energy is also dissipated at all scales but the amplitude of $D^c_{\eta}$ decreases much more faster in frequency as a consequence of a much more steep wave height power spectrum.

The theoretical frequency scaling of the dissipation spectrum of capillary wave is easily determined by combining the Kolmogorv-Zakharov solution of Eq.\ (\ref{wtcap}), $S_{\eta}\sim f^{-17/6}$, and the dissipation rate, from Eq.\ (\ref{eqS}), $\Gamma \sim f^{1/2}k$. Using the capillary wave dispersion relation $\omega^2\sim k^3$, we then obtain $D_{\eta}^c \sim S_{\eta} \Gamma \sim f^{-1/3}$. For low dissipation, we observe that $D_{\eta}^c$ is almost constant within the capillary inertial range ($f_{gc}<f<f_d$) as shown in on Fig. \ref{spectrediss} (top), and so in rough agreement with the $f^{-1/3}$ prediction. For high dissipation, $D_{\eta}^c$ is far from this theoretical scaling [see Fig. \ref{spectrediss} (bottom)], $S_{\eta}(f)$ being also much steeper than the theoretical wave spectrum (see inset).

\subsection{Dissipation at all scales}
In this part, we have experimentally determined the dissipated power in capillary and gravity waves and their corresponding spectra. We have shown quantitatively that dissipation occurs at all scales, and that only a small part of the power injected by the wave maker is linearly dissipated by waves. The main part must be dissipated either in the bulk, or by nonlinear wave dissipation processes (that are not taken into account in the present estimation of the wave dissipation) such as wave breakings \cite{MEL02,CHEN99} or the formation of capillary ripples on crested gravity waves  \cite{FM98,TSAI2010,Caulliez2013}.

We have also shown that the wave energy is mainly dissipated by gravity waves and that only a small part is transferred to capillary waves. In consequence, the capillary wave turbulence cascade is feed only by a small amount of the energy contained in the gravity waves, as already discussed in \cite{Deike2012}. Moreover, the dissipated power by capillary waves has been found to scale nonlinearly with $\epsilon_I$, at high enough injected power. Thus, the energy cascading through the capillary cascade is not proportional to the injected power. Instead of $\epsilon_I$, we will now define a quantity representing better the mean energy flux really cascading through the capillary scales.
 
\section{Estimation of the energy flux \label{part3}}
We will focus here to our experiments performed at {\it{low dissipation}} in which the frequency scaling of the experimental spectrum is found in agreement with the theoretical one of Eq.\ (\ref{wtcap}). Let us discuss now the spectrum scaling with the mean energy flux $\epsilon$. To do that, two experimental estimations of $\epsilon$ are used. 

\begin{figure}[t]
\begin{center}
\includegraphics[scale=0.45]{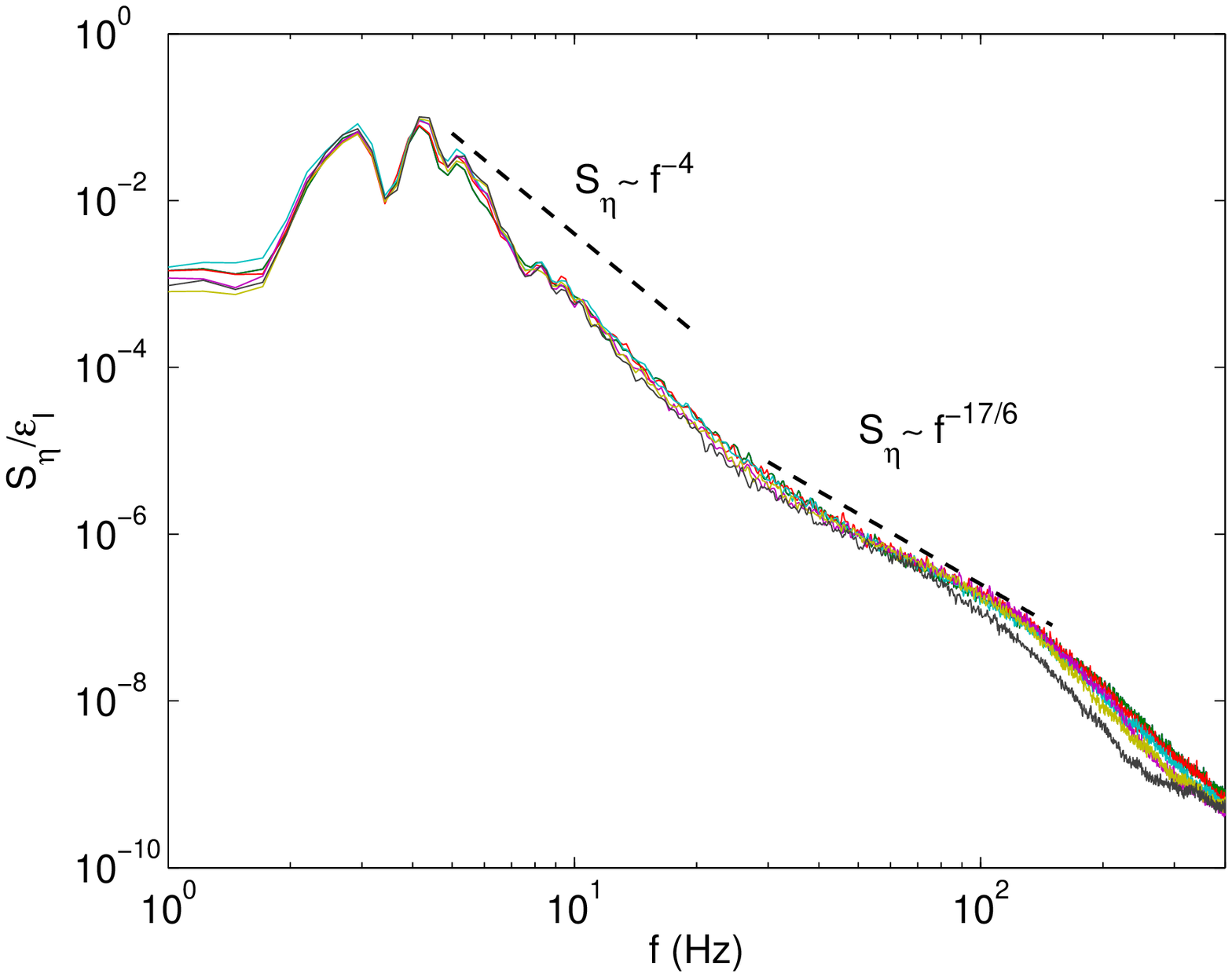}\\
\includegraphics[scale=0.45]{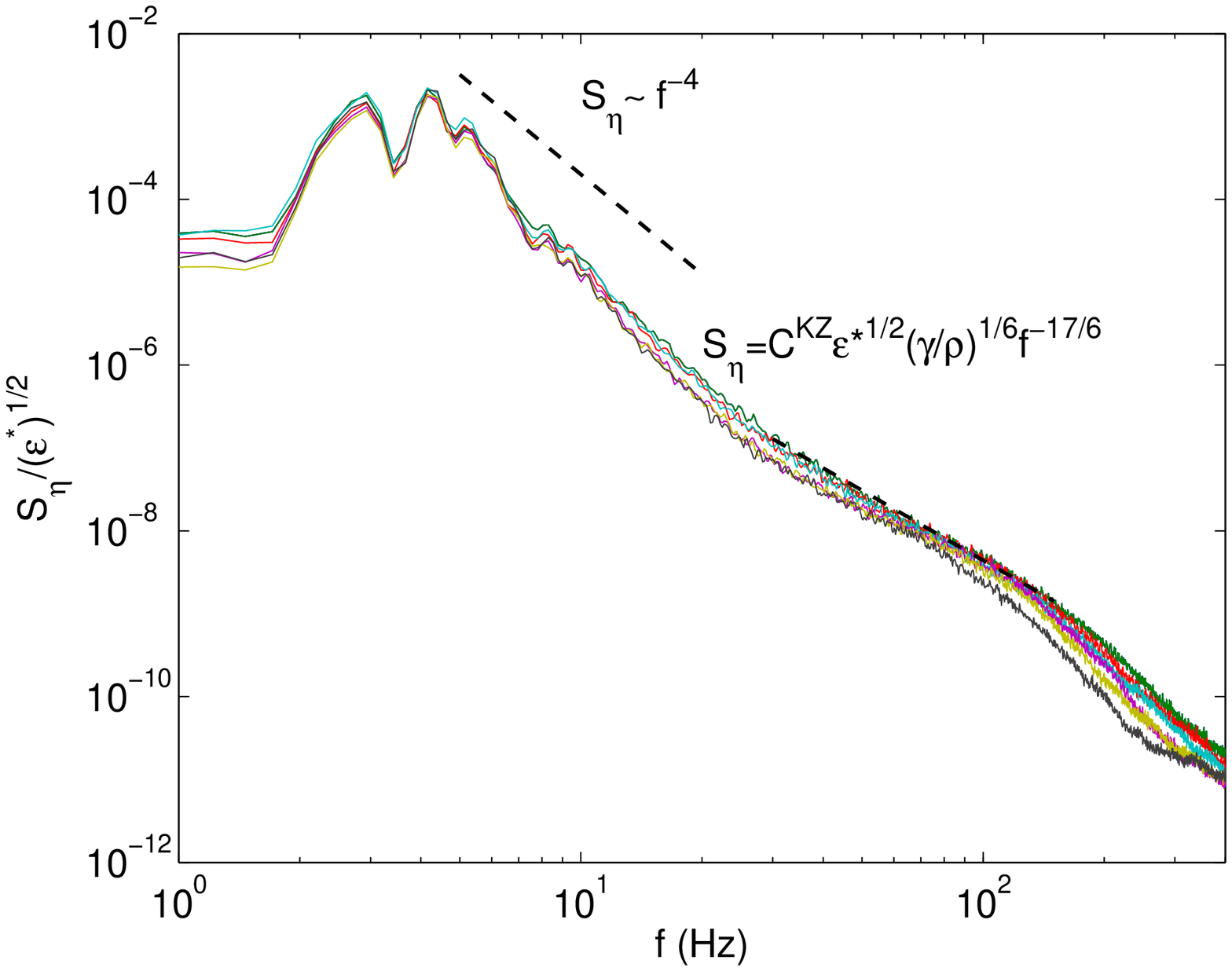}
\caption{(Color online) Rescaled power spectrum $S_{\eta}/\epsilon_I$ (top), and $S_{\eta}/{\epsilon^*}^{1/2}$ (bottom). In both cases, data roughly collapse on a single curve. Data are the same as in Fig. \ref{spvisc1}(top) ($10^{-4} \leq \epsilon_I \leq 5\ 10^{-4}$ m$^3$s$^{-3}$. Mercury). Dot-dashed line: theoretical capillary spectrum of Eq.\ (\ref{wtcap}) with $C^{KZ}=0.01$. Dashed lines: theoretical frequency scaling of the spectrum for gravity $S_{\eta}\sim f^{-4}$ and capillary $S_{\eta}\sim f^{-17/6}$ wave turbulence.}
\label{spectrenorm}
\end{center}
\end{figure}

First, $\epsilon$ is estimated straightforward by the mean injected power, $\epsilon\equiv \epsilon_I$, as previously proposed in \cite{Falcon07a,Xia10}. This estimation assumes that all the power injected into the system is injected into waves, then transferred through the gravity and capillary scales without dissipation, and finally dissipated at the end of the capillary cascade. The spectra of Fig. \ref{spvisc1}(top) normalized by $\epsilon_I^1$ displays a good collapse on a single curve as shown in Fig. \ref{spectrenorm}. However, as also reported previously \cite{Falcon07a,Xia10}, this $S_{\eta} \sim \epsilon_I^1$ scaling is in disagreement with the predicted one of Eq.\ (\ref{wtcap}) of weak turbulence theory. This discrepancy is explained by the presence of dissipation at all scales. Indeed, the mean injected power $\epsilon_I$ is not a good estimation of the energy flux within the capillary cascade since the energy dissipated by the capillary wave is not linearly dependent of $\epsilon_I$ as shown in \S \ref{part2}.

\begin{figure}[t]
\begin{center}
\includegraphics[scale=0.45]{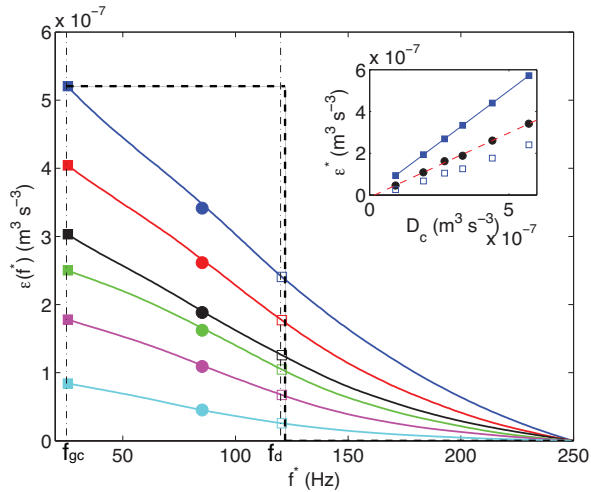}
\caption{(Color online) Experimental energy flux $\epsilon(f^*)$ at frequency $f^*$ estimated from Eq.\ (\ref{epsf}) for an increasing forcing amplitude (from bottom to top). Mean energy flux $\epsilon^*$ ($\bullet$) estimated from Eq.\ (\ref{meaneps}), energy flux at $f_{gc}$ [$\epsilon(f_{gc})=D_c$ ($\blacksquare$)] and at $f_d$ [$\epsilon(f_d)$ ($\square$)]. Same data as in Fig. \ref{spvisc1}(top). Vertical dot-dashed lines indicate $f_{gc}$ and $f_d$ delimiting the frequency range of the capillary cascade. Dashed lines: theoretical scenario of a constant flux in the inertial range and dissipation localized at $f_d$. Inset: Same symbols as in the main figure as a function of the dissipated power by capillary waves $D_c$. Dashed line is a linear fit. Solid line has a unit slope.}
\label{epsilonmes}
\end{center}
\end{figure}

A better way to estimate the energy flux is from the dissipated power by the capillary waves. The total power dissipated linearly by the capillary wave is given by Eq.\ (\ref{Dcap}), that is $D_c=\int_{f_{gc}}^{f_s/2}D_{\eta}^c(f)df$. This quantity integrates the power dissipated within the capillary cascade but also within the dissipative part of the spectrum. Thus, estimating $\epsilon\equiv D_c$ would lead to an overestimation of the mean energy flux. The power budget in the frequency Fourier space reads \cite{NazarenkoBook,Miquel2013}
\begin{equation}
\frac{\partial E_f}{\partial t}=-\frac{\partial \epsilon(f)}{\partial f}.
\end{equation} 
Consequently, the energy flux $\epsilon(f^*)$ at a given frequency $f^*$ reads
\begin{equation}
\epsilon(f^*)=\int_{f^*}^{f_s/2}D_{\eta}^c(f)df. \label{epsf}
\end{equation}
In practice, $\epsilon(f^*)$ is obtained using Eqs.\ (\ref{Detac}), (\ref{disseq}), (\ref{epsf}) and the experimental power spectrum of wave height, $S_{\eta}(f)$. Figure \ref{epsilonmes} shows $\epsilon(f^*)$ as a function of the frequency $f^*$ within the capillary range and for various forcing amplitudes. $\epsilon(f^*)$ is found to decreases with frequency since a part of energy is dissipated at each scale while another part is transferred to higher frequency. Thus, the theoretical scenario of weak turbulence where all the energy should be dissipated for frequencies larger than a critical dissipative frequency $f_d$ (see Fig. \ref{epsilonmes}) is not realistic in our experiments. A non constant energy flux through the scale has been also found numerically in wave turbulence on metallic plate in presence of dissipation at all scales \cite{Miquel2013}. 

The mean energy flux $\epsilon^*$ is then defined by the energy flux averaged through the capillary frequency range
\begin{equation}
\epsilon^*=\frac{\int_{f_{gc}}^{f_d}\epsilon(f)df }{f_d-f_{gc}}. \label{meaneps}
\end{equation}
Fig. \ref{epsilonmes} show the mean energy flux $\epsilon^*$ for different forcing amplitudes [see ($\bullet$) symbols]. These values roughly correspond to values of $\epsilon(f)$ at $f\approx80$ Hz in the middle of the cascade. The inset of Fig. \ref{epsilonmes} shows the evolution of $\epsilon^*$, the values of the flux at the beginning, $\epsilon(f_{gc})$, and at the end, $\epsilon(f_d)$, of the capillary cascade as a function of the dissipated power $D_c$ by capillary waves. Note that from Eq.\ (\ref{epsf}), Eq.\ (\ref{Dcap}) and Eq.\ (\ref{Detac}), one has $\epsilon(f_{gc})=D_c$. These three quantities depends linearly on $D_c$. Thus, rescaling the wave spectrum with one of these quantity would be equivalent. We choose $\epsilon \equiv \epsilon^*$ as an estimation of the energy flux cascading through the capillary scales. Figure \ref{spectrenorm} (bottom) then shows the rescaled spectrum $S_{\eta}/(\epsilon^*)^{1/2}$ where all curves roughly collapse on a single curve.

To sum up, we have shown that dissipation at all scales explains the previous controversy of the scaling of the capillary wave spectrum with the mean energy flux. A new estimation of the flux has been proposed from the dissipated power. The energy flux is then found to be non constant over the scales. Nevertheless, the estimation of the mean energy flux allows us to rescale properly the wave height spectrum, and we observe $S_{\eta}(f)\sim \epsilon^{1/2}f^{-17/6}$ in agreement with the theory of capillary wave turbulence.

\section{Estimation of the Kolmogorov-Zakharov constant\label{part4}}
In the previous section, we have shown that both the scalings of the capillary spectrum with frequency and with the mean energy flux $\epsilon^*$ are found in agreement with wave turbulence theory of Eq.\ (\ref{wtcap}). We can thus now evaluate experimentally the Kolmogorov Zakharov constant $C^{KZ}$ using the estimation $\epsilon\equiv \epsilon^*$. Fig. \ref{spectrenorm}(bottom) shows the Kolmogorov-Zakharov spectrum $S_{\eta}(f)=C_{exp}^{KZ} {\epsilon^{*}}^{1/2}\left(\frac{\gamma}{\rho}\right)^{1/6}f^{-17/6}$, where the constant $C_{exp}^{KZ}$ is the experimentally fitted (see dot-dashed line). One finds $C_{exp}^{KZ} \approx 0.01$.

The $C^{KZ}$ constant was previsouly calculated from the wave action spectrum $n_k$ \cite{Pushkarev2000}. The relation between the constants defined from $n_k$ ($C_{n_{k}}^{KZ}$), and from $S_{\eta}(f)$ ($C^{KZ}$) is given by $C^{KZ}=\frac{4\pi}{3} C_{n_{k}}^{KZ}(2\pi)^{-17/6}$. The $(2\pi)^{-17/6}$ factor is due to the change from frequency $f$ to the pulsation $\omega$, and the $\frac{4\pi}{3}$ factor comes from the relation between $n_k$ and $S_{\eta}(f)$. Theoretically, $C_{n_{k}}^{th}=9.85$ \cite{Pushkarev2000}, while in the numerical simulations, $C_{n_{k}}^{n}\approx 1.7$ \cite{Pushkarev2000}. The difference between theory and numerics is explained by the small inertial range and the existence of numerical dissipation \cite{Pushkarev2000}. 

Here, we experimentally found $C^{KZ}\approx 0.01$ that corresponds to $C_{n_{k}}^{KZ}\approx 0.5$. This value is 3.4 times smaller than the numerical value, and 20 times smaller that the theoretical one. However, we have to keep in mind that dissipation occurs at all scales experimentally, and that a non constant energy flux through the scales is observed contrary to the theoretical hypotheses.

\section{Conclusion\label{Conclusion}}
In this paper, we have discussed the influence of dissipation on gravity-capillary wave turbulence. We have shown that the main part of the injected energy at large scale is dissipated by gravity waves and only a small part of it is transferred to capillary waves. This show that evaluating the energy flux by the mean injected power is not a valid approximation. We propose an estimation of the energy flux within the capillary cascade, related to the linear dissipated power by the capillary waves with the cascade inertial range.

A capillary wave turbulence regime with a wave spectrum as a power law of the scale is observed whatever the intensity of the dissipation but two regimes can be defined, depending on the level of dissipation in the system.

When the dissipation is low enough, the wave spectrum is found in good agreement in regards to the frequency scaling and the energy flux scaling (newly defined). This result explains the previous controversy on the energy flux scaling of the capillary wave spectrum \cite{Falcon07a,Xia10}, pointed out as an open question in a recent review \cite{NewellReview}. The Kolmogorov-Zakharov constant is then evaluated experimentally, for the first time. The value is found one order of magnitude smaller than the one predicted by the theory, since dissipation occurring at all scales is observed experimentally, as well as a non constant energy flux through the scales contrary to the theoretical hypotheses.

When the dissipation goes beyond a certain threshold, the power law spectrum becomes steeper and the agreement with the theory is lost. The spectrum becomes steeper and steeper when the dissipation is further increased. This latter has also been observed experimentally and numerically in flexural wave turbulence \cite{Humbert2013,Miquel2013}. It is possible that dissipation is also responsible for the discrepancy between theory and experiment observed in wave turbulence at the surface of floating elastic sheet \cite{Deike2013}. Moreover, at high dissipation, the capillary wave spectrum is found to depend on the injected power, which reminds us results in gravity wave turbulence \cite{Falcon07a,Denissenko07,Nazarenko2010}. Thus dissipation appears to be of prior importance to explain the differences between weak turbulence theory and experimental wave turbulence regimes.

The next step would to explain quantitatively the threshold from the {\it{low dissipation}} to the {\it{high dissipation}} situations. The measurement of the non linear interaction time $\tau_{nl}$ and its comparison with the dissipation time should be the starting point. Experimentally, a direct measurement of the energy flux in the $k$-space (in a similar way to what is done numerically \cite{Miquel2013}) remains an important challenge and would be of interest to confirm our results and discuss the non linear time. Moreover, it would be interesting to be able to close the power budget, by means of surface and bulk measurements. A better understanding of non linear dissipation processes appears also necessary, as wave breaking and the occurrence of ripples on gravity waves. The inclusion of these kind of coherent structures, as well as the coexistence of dissipation and energy transfers appears as important challenges to improve our understanding of natural wave turbulence system.

\begin{acknowledgments}
We thank C. Laroche for technical help and S. Fauve for discussions. This work has been supported by ANR Turbulon 12-BS04-0005.
\end{acknowledgments}
\bibliographystyle{apsrev4-1}
\bibliography{wtdiss2.bib}

\begin{thebibliography}{34}%
\makeatletter
\providecommand \@ifxundefined [1]{%
 \@ifx{#1\undefined}
}%
\providecommand \@ifnum [1]{%
 \ifnum #1\expandafter \@firstoftwo
 \else \expandafter \@secondoftwo
 \fi
}%
\providecommand \@ifx [1]{%
 \ifx #1\expandafter \@firstoftwo
 \else \expandafter \@secondoftwo
 \fi
}%
\providecommand \natexlab [1]{#1}%
\providecommand \enquote  [1]{``#1''}%
\providecommand \bibnamefont  [1]{#1}%
\providecommand \bibfnamefont [1]{#1}%
\providecommand \citenamefont [1]{#1}%
\providecommand \href@noop [0]{\@secondoftwo}%
\providecommand \href [0]{\begingroup \@sanitize@url \@href}%
\providecommand \@href[1]{\@@startlink{#1}\@@href}%
\providecommand \@@href[1]{\endgroup#1\@@endlink}%
\providecommand \@sanitize@url [0]{\catcode `\\12\catcode `\$12\catcode
  `\&12\catcode `\#12\catcode `\^12\catcode `\_12\catcode `\%12\relax}%
\providecommand \@@startlink[1]{}%
\providecommand \@@endlink[0]{}%
\providecommand \url  [0]{\begingroup\@sanitize@url \@url }%
\providecommand \@url [1]{\endgroup\@href {#1}{\urlprefix }}%
\providecommand \urlprefix  [0]{URL }%
\providecommand \Eprint [0]{\href }%
\providecommand \doibase [0]{http://dx.doi.org/}%
\providecommand \selectlanguage [0]{\@gobble}%
\providecommand \bibinfo  [0]{\@secondoftwo}%
\providecommand \bibfield  [0]{\@secondoftwo}%
\providecommand \translation [1]{[#1]}%
\providecommand \BibitemOpen [0]{}%
\providecommand \bibitemStop [0]{}%
\providecommand \bibitemNoStop [0]{.\EOS\space}%
\providecommand \EOS [0]{\spacefactor3000\relax}%
\providecommand \BibitemShut  [1]{\csname bibitem#1\endcsname}%
\let\auto@bib@innerbib\@empty
\bibitem [{\citenamefont {Zakharov}\ \emph {et~al.}(1992)\citenamefont
  {Zakharov}, \citenamefont {Falkovitch},\ and\ \citenamefont
  {L'vov.}}]{ZakharovBook}%
  \BibitemOpen
  \bibfield  {author} {\bibinfo {author} {\bibfnamefont {V.~E.}\ \bibnamefont
  {Zakharov}}, \bibinfo {author} {\bibfnamefont {G.}~\bibnamefont
  {Falkovitch}}, \ and\ \bibinfo {author} {\bibfnamefont {V.~S.}\ \bibnamefont
  {L'vov.}},\ }\href@noop {} {\emph {\bibinfo {title} {Kolmogorov spectra of
  turbulence. I. Wave turbulence}}}\ (\bibinfo  {publisher} {Springer-Verlag},\
  \bibinfo {year} {1992})\BibitemShut {NoStop}%
\bibitem [{\citenamefont {Nazarenko}(2011)}]{NazarenkoBook}%
  \BibitemOpen
  \bibfield  {author} {\bibinfo {author} {\bibfnamefont {S.}~\bibnamefont
  {Nazarenko}},\ }\href@noop {} {\emph {\bibinfo {title} {Wave turbulence}}}\
  (\bibinfo  {publisher} {Springer-Verlag},\ \bibinfo {year}
  {2011})\BibitemShut {NoStop}%
\bibitem [{\citenamefont {Falcon}(2010)}]{FalconReview}%
  \BibitemOpen
  \bibfield  {author} {\bibinfo {author} {\bibfnamefont {E.}~\bibnamefont
  {Falcon}},\ }\href@noop {} {\bibfield  {journal} {\bibinfo  {journal}
  {Discret. Contin. Dyn. S. B}\ }\textbf {\bibinfo {volume} {13}},\ \bibinfo
  {pages} {819} (\bibinfo {year} {2010})}\BibitemShut {NoStop}%
\bibitem [{\citenamefont {Newell}\ and\ \citenamefont
  {Rumpf}(2011)}]{NewellReview}%
  \BibitemOpen
  \bibfield  {author} {\bibinfo {author} {\bibfnamefont {A.~C.}\ \bibnamefont
  {Newell}}\ and\ \bibinfo {author} {\bibfnamefont {B.}~\bibnamefont {Rumpf}},\
  }\href@noop {} {\bibfield  {journal} {\bibinfo  {journal} {An. Rev. Fluid
  Mech.}\ }\textbf {\bibinfo {volume} {43}},\ \bibinfo {pages} {59} (\bibinfo
  {year} {2011})}\BibitemShut {NoStop}%
\bibitem [{\citenamefont {Humbert}\ \emph {et~al.}(2013)\citenamefont
  {Humbert}, \citenamefont {Cadot}, \citenamefont {During}, \citenamefont
  {Josserand}, \citenamefont {Rica},\ and\ \citenamefont
  {Touze}}]{Humbert2013}%
  \BibitemOpen
  \bibfield  {author} {\bibinfo {author} {\bibfnamefont {T.}~\bibnamefont
  {Humbert}}, \bibinfo {author} {\bibfnamefont {O.}~\bibnamefont {Cadot}},
  \bibinfo {author} {\bibfnamefont {G.}~\bibnamefont {During}}, \bibinfo
  {author} {\bibfnamefont {C.}~\bibnamefont {Josserand}}, \bibinfo {author}
  {\bibfnamefont {S.}~\bibnamefont {Rica}}, \ and\ \bibinfo {author}
  {\bibfnamefont {C.}~\bibnamefont {Touze}},\ }\href@noop {} {\bibfield
  {journal} {\bibinfo  {journal} {EPL}\ }\textbf {\bibinfo {volume} {102}},\
  \bibinfo {pages} {30002} (\bibinfo {year} {2013})}\BibitemShut {NoStop}%
\bibitem [{\citenamefont {During}\ \emph {et~al.}(2006)\citenamefont {During},
  \citenamefont {Josserand},\ and\ \citenamefont {Rica}}]{Josserand}%
  \BibitemOpen
  \bibfield  {author} {\bibinfo {author} {\bibfnamefont {G.}~\bibnamefont
  {During}}, \bibinfo {author} {\bibfnamefont {C.}~\bibnamefont {Josserand}}, \
  and\ \bibinfo {author} {\bibfnamefont {S.}~\bibnamefont {Rica}},\ }\href
  {\doibase 10.1103/PhysRevLett.97.025503} {\bibfield  {journal} {\bibinfo
  {journal} {Phys. Rev. Lett.}\ }\textbf {\bibinfo {volume} {97}},\ \bibinfo
  {pages} {025503} (\bibinfo {year} {2006})}\BibitemShut {NoStop}%
\bibitem [{\citenamefont {Miquel}\ \emph {et~al.}(2013)\citenamefont {Miquel},
  \citenamefont {Alexakis},\ and\ \citenamefont {Mordant}}]{Miquel2013}%
  \BibitemOpen
  \bibfield  {author} {\bibinfo {author} {\bibfnamefont {B.}~\bibnamefont
  {Miquel}}, \bibinfo {author} {\bibfnamefont {A.}~\bibnamefont {Alexakis}}, \
  and\ \bibinfo {author} {\bibfnamefont {N.}~\bibnamefont {Mordant}},\
  }\href@noop {} {\bibfield  {journal} {\bibinfo  {journal} {submitted to Phys.
  Rev. E}\ } (\bibinfo {year} {2013})}\BibitemShut {NoStop}%
\bibitem [{\citenamefont {Falcon}\ \emph {et~al.}(2007)\citenamefont {Falcon},
  \citenamefont {Laroche},\ and\ \citenamefont {Fauve}}]{Falcon07a}%
  \BibitemOpen
  \bibfield  {author} {\bibinfo {author} {\bibfnamefont {E.}~\bibnamefont
  {Falcon}}, \bibinfo {author} {\bibfnamefont {C.}~\bibnamefont {Laroche}}, \
  and\ \bibinfo {author} {\bibfnamefont {S.}~\bibnamefont {Fauve}},\
  }\href@noop {} {\bibfield  {journal} {\bibinfo  {journal} {Phys. Rev. Lett.}\
  }\textbf {\bibinfo {volume} {{98}}},\ \bibinfo {pages} {{094503}} (\bibinfo
  {year} {{2007}})}\BibitemShut {NoStop}%
\bibitem [{\citenamefont {Berhanu}\ and\ \citenamefont
  {Falcon}(2013)}]{Berhanu2013}%
  \BibitemOpen
  \bibfield  {author} {\bibinfo {author} {\bibfnamefont {M.}~\bibnamefont
  {Berhanu}}\ and\ \bibinfo {author} {\bibfnamefont {E.}~\bibnamefont
  {Falcon}},\ }\href {\doibase 10.1103/PhysRevE.87.033003} {\bibfield
  {journal} {\bibinfo  {journal} {Phys. Rev. E}\ }\textbf {\bibinfo {volume}
  {87}},\ \bibinfo {pages} {033003} (\bibinfo {year} {2013})}\BibitemShut
  {NoStop}%
\bibitem [{\citenamefont {Issenmann}\ and\ \citenamefont
  {Falcon}(2013)}]{Issenmann}%
  \BibitemOpen
  \bibfield  {author} {\bibinfo {author} {\bibfnamefont {B.}~\bibnamefont
  {Issenmann}}\ and\ \bibinfo {author} {\bibfnamefont {E.}~\bibnamefont
  {Falcon}},\ }\href {\doibase 10.1103/PhysRevE.87.011001} {\bibfield
  {journal} {\bibinfo  {journal} {Phys. Rev. E}\ }\textbf {\bibinfo {volume}
  {87}},\ \bibinfo {pages} {011001} (\bibinfo {year} {2013})}\BibitemShut
  {NoStop}%
\bibitem [{\citenamefont {Falc{\'o}n}\ \emph {et~al.}(2009)\citenamefont
  {Falc{\'o}n}, \citenamefont {Falcon}, \citenamefont {Bortolozzo},\ and\
  \citenamefont {Fauve}}]{Falcon0g}%
  \BibitemOpen
  \bibfield  {author} {\bibinfo {author} {\bibfnamefont {C.}~\bibnamefont
  {Falc{\'o}n}}, \bibinfo {author} {\bibfnamefont {E.}~\bibnamefont {Falcon}},
  \bibinfo {author} {\bibfnamefont {U.}~\bibnamefont {Bortolozzo}}, \ and\
  \bibinfo {author} {\bibfnamefont {S.}~\bibnamefont {Fauve}},\ }\href
  {http://stacks.iop.org/0295-5075/86/i=1/a=14002} {\bibfield  {journal}
  {\bibinfo  {journal} {EPL (Europhysics Letters)}\ }\textbf {\bibinfo {volume}
  {86}},\ \bibinfo {pages} {14002} (\bibinfo {year} {2009})}\BibitemShut
  {NoStop}%
\bibitem [{\citenamefont {Holt}\ and\ \citenamefont {Trinh}(1996)}]{Holt1996}%
  \BibitemOpen
  \bibfield  {author} {\bibinfo {author} {\bibfnamefont {R.~G.}\ \bibnamefont
  {Holt}}\ and\ \bibinfo {author} {\bibfnamefont {E.~H.}\ \bibnamefont
  {Trinh}},\ }\href {\doibase 10.1103/PhysRevLett.77.1274} {\bibfield
  {journal} {\bibinfo  {journal} {Phys. Rev. Lett.}\ }\textbf {\bibinfo
  {volume} {77}},\ \bibinfo {pages} {1274} (\bibinfo {year}
  {1996})}\BibitemShut {NoStop}%
\bibitem [{\citenamefont {Wright}\ \emph {et~al.}(1996)\citenamefont {Wright},
  \citenamefont {Budakian},\ and\ \citenamefont {Putterman}}]{Putterman1996}%
  \BibitemOpen
  \bibfield  {author} {\bibinfo {author} {\bibfnamefont {W.~B.}\ \bibnamefont
  {Wright}}, \bibinfo {author} {\bibfnamefont {R.}~\bibnamefont {Budakian}}, \
  and\ \bibinfo {author} {\bibfnamefont {S.~J.}\ \bibnamefont {Putterman}},\
  }\href@noop {} {\bibfield  {journal} {\bibinfo  {journal} {Phys. Rev. Lett.}\
  }\textbf {\bibinfo {volume} {76}},\ \bibinfo {pages} {4528} (\bibinfo {year}
  {1996})}\BibitemShut {NoStop}%
\bibitem [{\citenamefont {Henry}\ \emph {et~al.}(2000)\citenamefont {Henry},
  \citenamefont {Alstrom},\ and\ \citenamefont {Levinsen}}]{Levinsen2000}%
  \BibitemOpen
  \bibfield  {author} {\bibinfo {author} {\bibfnamefont {E.}~\bibnamefont
  {Henry}}, \bibinfo {author} {\bibfnamefont {P.}~\bibnamefont {Alstrom}}, \
  and\ \bibinfo {author} {\bibfnamefont {M.~T.}\ \bibnamefont {Levinsen}},\
  }\href {http://stacks.iop.org/0295-5075/52/i=1/a=027} {\bibfield  {journal}
  {\bibinfo  {journal} {EPL}\ }\textbf {\bibinfo {volume} {52}},\ \bibinfo
  {pages} {27} (\bibinfo {year} {2000})}\BibitemShut {NoStop}%
\bibitem [{\citenamefont {Brazhnikov}\ \emph {et~al.}(2002)\citenamefont
  {Brazhnikov}, \citenamefont {Kolmakov},\ and\ \citenamefont
  {Levchenko}}]{Brazhnikov02}%
  \BibitemOpen
  \bibfield  {author} {\bibinfo {author} {\bibfnamefont {M.}~\bibnamefont
  {Brazhnikov}}, \bibinfo {author} {\bibfnamefont {G.}~\bibnamefont
  {Kolmakov}}, \ and\ \bibinfo {author} {\bibfnamefont {A.}~\bibnamefont
  {Levchenko}},\ }\href@noop {} {\bibfield  {journal} {\bibinfo  {journal}
  {Journal of Experimental and Theoretical Physics}\ }\textbf {\bibinfo
  {volume} {95}},\ \bibinfo {pages} {447} (\bibinfo {year} {2002})}\BibitemShut
  {NoStop}%
\bibitem [{\citenamefont {Xia}\ \emph {et~al.}(2010)\citenamefont {Xia},
  \citenamefont {Shats},\ and\ \citenamefont {Punzmann}}]{Xia10}%
  \BibitemOpen
  \bibfield  {author} {\bibinfo {author} {\bibfnamefont {H.}~\bibnamefont
  {Xia}}, \bibinfo {author} {\bibfnamefont {M.}~\bibnamefont {Shats}}, \ and\
  \bibinfo {author} {\bibfnamefont {H.}~\bibnamefont {Punzmann}},\ }\href@noop
  {} {\bibfield  {journal} {\bibinfo  {journal} {EPL}\ }\textbf {\bibinfo
  {volume} {91}},\ \bibinfo {pages} {14002} (\bibinfo {year}
  {2010})}\BibitemShut {NoStop}%
\bibitem [{\citenamefont {Pushkarev}\ and\ \citenamefont
  {Zakharov}(1996)}]{Pushkarev1996}%
  \BibitemOpen
  \bibfield  {author} {\bibinfo {author} {\bibfnamefont {A.~N.}\ \bibnamefont
  {Pushkarev}}\ and\ \bibinfo {author} {\bibfnamefont {V.~E.}\ \bibnamefont
  {Zakharov}},\ }\href {\doibase 10.1103/PhysRevLett.76.3320} {\bibfield
  {journal} {\bibinfo  {journal} {Phys. Rev. Lett.}\ }\textbf {\bibinfo
  {volume} {76}},\ \bibinfo {pages} {3320} (\bibinfo {year}
  {1996})}\BibitemShut {NoStop}%
\bibitem [{\citenamefont {Pushkarev}\ and\ \citenamefont
  {Zakharov}(2000)}]{Pushkarev2000}%
  \BibitemOpen
  \bibfield  {author} {\bibinfo {author} {\bibfnamefont {A.}~\bibnamefont
  {Pushkarev}}\ and\ \bibinfo {author} {\bibfnamefont {V.}~\bibnamefont
  {Zakharov}},\ }\href {\doibase 10.1016/S0167-2789(99)00069-X} {\bibfield
  {journal} {\bibinfo  {journal} {Physica D: Nonlinear Phenomena}\ }\textbf
  {\bibinfo {volume} {135}},\ \bibinfo {pages} {98 } (\bibinfo {year}
  {2000})}\BibitemShut {NoStop}%
\bibitem [{\citenamefont {Denissenko}\ \emph {et~al.}(2007)\citenamefont
  {Denissenko}, \citenamefont {Lukaschuk},\ and\ \citenamefont
  {Nazarenko}}]{Denissenko07}%
  \BibitemOpen
  \bibfield  {author} {\bibinfo {author} {\bibfnamefont {P.}~\bibnamefont
  {Denissenko}}, \bibinfo {author} {\bibfnamefont {S.}~\bibnamefont
  {Lukaschuk}}, \ and\ \bibinfo {author} {\bibfnamefont {S.}~\bibnamefont
  {Nazarenko}},\ }\href@noop {} {\bibfield  {journal} {\bibinfo  {journal}
  {Phys. Rev. Lett.}\ }\textbf {\bibinfo {volume} {99}},\ \bibinfo {pages}
  {014501} (\bibinfo {year} {2007})}\BibitemShut {NoStop}%
\bibitem [{\citenamefont {Deike}\ \emph {et~al.}(2012)\citenamefont {Deike},
  \citenamefont {Berhanu},\ and\ \citenamefont {Falcon}}]{Deike2012}%
  \BibitemOpen
  \bibfield  {author} {\bibinfo {author} {\bibfnamefont {L.}~\bibnamefont
  {Deike}}, \bibinfo {author} {\bibfnamefont {M.}~\bibnamefont {Berhanu}}, \
  and\ \bibinfo {author} {\bibfnamefont {E.}~\bibnamefont {Falcon}},\
  }\href@noop {} {\bibfield  {journal} {\bibinfo  {journal} {Phys. Rev. E}\
  }\textbf {\bibinfo {volume} {85}},\ \bibinfo {pages} {066311} (\bibinfo
  {year} {2012})}\BibitemShut {NoStop}%
\bibitem [{\citenamefont {Miles}(1967)}]{Miles}%
  \BibitemOpen
  \bibfield  {author} {\bibinfo {author} {\bibfnamefont {J.~W.}\ \bibnamefont
  {Miles}},\ }\href@noop {} {\bibfield  {journal} {\bibinfo  {journal}
  {Proceedings of the Royal Society of London. Series A, Mathematical and
  Physical Sciences}\ }\textbf {\bibinfo {volume} {297}},\ \bibinfo {pages}
  {pp. 459} (\bibinfo {year} {1967})}\BibitemShut {NoStop}%
\bibitem [{\citenamefont {Landau}\ and\ \citenamefont
  {Lifchitz}(1951)}]{LandauFluid}%
  \BibitemOpen
  \bibfield  {author} {\bibinfo {author} {\bibfnamefont {L.}~\bibnamefont
  {Landau}}\ and\ \bibinfo {author} {\bibfnamefont {F.}~\bibnamefont
  {Lifchitz}},\ }\href@noop {} {\emph {\bibinfo {title} {Mecanique des
  fluides}}}\ (\bibinfo  {publisher} {Editions Mir},\ \bibinfo {year}
  {1951})\BibitemShut {NoStop}%
\bibitem [{\citenamefont {Lamb}(1932)}]{Lamb}%
  \BibitemOpen
  \bibfield  {author} {\bibinfo {author} {\bibfnamefont {H.}~\bibnamefont
  {Lamb}},\ }\href@noop {} {\emph {\bibinfo {title} {Hydrodynamics}}}\
  (\bibinfo  {publisher} {Dover},\ \bibinfo {year} {1932})\BibitemShut
  {NoStop}%
\bibitem [{\citenamefont {Dorn}(1966)}]{VanDorn}%
  \BibitemOpen
  \bibfield  {author} {\bibinfo {author} {\bibfnamefont {W.~G.~V.}\
  \bibnamefont {Dorn}},\ }\href@noop {} {\bibfield  {journal} {\bibinfo
  {journal} {J. Fluid Mech.}\ }\textbf {\bibinfo {volume} {24}},\ \bibinfo
  {pages} {pp 769} (\bibinfo {year} {1966})}\BibitemShut {NoStop}%
\bibitem [{\citenamefont {Henderson}\ and\ \citenamefont
  {Miles}(1990)}]{HendersonMiles}%
  \BibitemOpen
  \bibfield  {author} {\bibinfo {author} {\bibfnamefont {D.~M.}\ \bibnamefont
  {Henderson}}\ and\ \bibinfo {author} {\bibfnamefont {J.~W.}\ \bibnamefont
  {Miles}},\ }\href@noop {} {\bibfield  {journal} {\bibinfo  {journal} {J.
  Fluid Mech.}\ }\textbf {\bibinfo {volume} {213}},\ \bibinfo {pages} {95}
  (\bibinfo {year} {1990})}\BibitemShut {NoStop}%
\bibitem [{\citenamefont {Association}(1963)}]{Glycerin}%
  \BibitemOpen
  \bibfield  {author} {\bibinfo {author} {\bibfnamefont {G.~P.}\ \bibnamefont
  {Association}},\ }\href@noop {} {\emph {\bibinfo {title} {Physical properties
  of glycerine and its solutions.}}}\ (\bibinfo  {publisher} {New York :
  Glycerine Producers' Association},\ \bibinfo {year} {1963})\BibitemShut
  {NoStop}%
\bibitem [{\citenamefont {Falcon}\ \emph {et~al.}(2008)\citenamefont {Falcon},
  \citenamefont {Aumaitre}, \citenamefont {Falcon}, \citenamefont {Laroche},\
  and\ \citenamefont {Fauve}}]{FalconFlux}%
  \BibitemOpen
  \bibfield  {author} {\bibinfo {author} {\bibfnamefont {E.}~\bibnamefont
  {Falcon}}, \bibinfo {author} {\bibfnamefont {S.}~\bibnamefont {Aumaitre}},
  \bibinfo {author} {\bibfnamefont {C.}~\bibnamefont {Falcon}}, \bibinfo
  {author} {\bibfnamefont {C.}~\bibnamefont {Laroche}}, \ and\ \bibinfo
  {author} {\bibfnamefont {S.}~\bibnamefont {Fauve}},\ }\href {\doibase
  10.1103/PhysRevLett.100.064503} {\bibfield  {journal} {\bibinfo  {journal}
  {Phys. Rev. Lett.}\ }\textbf {\bibinfo {volume} {100}},\ \bibinfo {pages}
  {064503} (\bibinfo {year} {2008})}\BibitemShut {NoStop}%
\bibitem [{\citenamefont {Nazarenko}\ \emph {et~al.}(2010)\citenamefont
  {Nazarenko}, \citenamefont {Lukashuk}, \citenamefont {McLelland},\ and\
  \citenamefont {Denissenko}}]{Nazarenko2010}%
  \BibitemOpen
  \bibfield  {author} {\bibinfo {author} {\bibfnamefont {S.}~\bibnamefont
  {Nazarenko}}, \bibinfo {author} {\bibfnamefont {S.}~\bibnamefont {Lukashuk}},
  \bibinfo {author} {\bibfnamefont {S.}~\bibnamefont {McLelland}}, \ and\
  \bibinfo {author} {\bibfnamefont {P.}~\bibnamefont {Denissenko}},\ }\href
  {\doibase 10.1017/S0022112009991820} {\bibfield  {journal} {\bibinfo
  {journal} {J. Fluid Mech.}\ }\textbf {\bibinfo {volume} {642}},\ \bibinfo
  {pages} {395} (\bibinfo {year} {2010})}\BibitemShut {NoStop}%
\bibitem [{\citenamefont {Deike}\ \emph {et~al.}(2013)\citenamefont {Deike},
  \citenamefont {Bacri},\ and\ \citenamefont {Falcon}}]{Deike2013}%
  \BibitemOpen
  \bibfield  {author} {\bibinfo {author} {\bibfnamefont {L.}~\bibnamefont
  {Deike}}, \bibinfo {author} {\bibfnamefont {J.-C.}\ \bibnamefont {Bacri}}, \
  and\ \bibinfo {author} {\bibfnamefont {E.}~\bibnamefont {Falcon}},\
  }\href@noop {} {\bibfield  {journal} {\bibinfo  {journal} {J. Fluid Mech.}\
  }\textbf {\bibinfo {volume} {in press}} (\bibinfo {year} {2013})}\BibitemShut
  {NoStop}%
\bibitem [{\citenamefont {Melville}\ \emph {et~al.}(2002)\citenamefont
  {Melville}, \citenamefont {Veron},\ and\ \citenamefont {White}}]{MEL02}%
  \BibitemOpen
  \bibfield  {author} {\bibinfo {author} {\bibfnamefont {W.~K.}\ \bibnamefont
  {Melville}}, \bibinfo {author} {\bibfnamefont {F.}~\bibnamefont {Veron}}, \
  and\ \bibinfo {author} {\bibfnamefont {C.~J.}\ \bibnamefont {White}},\
  }\href@noop {} {\bibfield  {journal} {\bibinfo  {journal} {J. Fluid. Mech.}\
  }\textbf {\bibinfo {volume} {454}} (\bibinfo {year} {2002})}\BibitemShut
  {NoStop}%
\bibitem [{\citenamefont {Chen}\ \emph {et~al.}(1999)\citenamefont {Chen},
  \citenamefont {Kharif}, \citenamefont {Zaleski},\ and\ \citenamefont
  {Li}}]{CHEN99}%
  \BibitemOpen
  \bibfield  {author} {\bibinfo {author} {\bibfnamefont {G.}~\bibnamefont
  {Chen}}, \bibinfo {author} {\bibfnamefont {C.}~\bibnamefont {Kharif}},
  \bibinfo {author} {\bibfnamefont {S.}~\bibnamefont {Zaleski}}, \ and\
  \bibinfo {author} {\bibfnamefont {J.}~\bibnamefont {Li}},\ }\href@noop {}
  {\bibfield  {journal} {\bibinfo  {journal} {Physics of fluid}\ }\textbf
  {\bibinfo {volume} {11}},\ \bibinfo {pages} {121} (\bibinfo {year}
  {1999})}\BibitemShut {NoStop}%
\bibitem [{\citenamefont {Fedorov}\ and\ \citenamefont
  {Melville}(1998)}]{FM98}%
  \BibitemOpen
  \bibfield  {author} {\bibinfo {author} {\bibfnamefont {A.~V.}\ \bibnamefont
  {Fedorov}}\ and\ \bibinfo {author} {\bibfnamefont {W.~K.}\ \bibnamefont
  {Melville}},\ }\href@noop {} {\bibfield  {journal} {\bibinfo  {journal}
  {J.Fluid Mech}\ }\textbf {\bibinfo {volume} {354}},\ \bibinfo {pages} {1}
  (\bibinfo {year} {1998})}\BibitemShut {NoStop}%
\bibitem [{\citenamefont {Tsai}\ and\ \citenamefont {ping
  Hung}(2010)}]{TSAI2010}%
  \BibitemOpen
  \bibfield  {author} {\bibinfo {author} {\bibfnamefont {W.-t.}\ \bibnamefont
  {Tsai}}\ and\ \bibinfo {author} {\bibfnamefont {L.}~\bibnamefont {ping
  Hung}},\ }\href@noop {} {\bibfield  {journal} {\bibinfo  {journal} {J. Phys.
  Oceanogr.}\ }\textbf {\bibinfo {volume} {40}},\ \bibinfo {pages} {2435}
  (\bibinfo {year} {2010})}\BibitemShut {NoStop}%
\bibitem [{\citenamefont {Caulliez}(2013)}]{Caulliez2013}%
  \BibitemOpen
  \bibfield  {author} {\bibinfo {author} {\bibfnamefont {G.}~\bibnamefont
  {Caulliez}},\ }\href {\doibase 10.1029/2012JC008402} {\bibfield  {journal}
  {\bibinfo  {journal} {Journal of Geophysical Research: Oceans}\ }\textbf
  {\bibinfo {volume} {118}},\ \bibinfo {pages} {672} (\bibinfo {year}
  {2013})}\BibitemShut {NoStop}%
\end{thebibliography}%
\end{document}